# Bursty and persistent properties of large-scale brain networks revealed with a point-based method for dynamic functional connectivity.

William Hedley Thompson[1] & Peter Fransson[1]

[1]Department of Clinical Neuroscience, Karolinska Institutet, Stockholm, Sweden
E-mail: whthompson@ymail.com; Peter.Fransson@ki.se

13-January-2016

## Abstract

In this paper, we present a novel and versatile method to study the dynamics of resting-state fMRI brain connectivity with a high temporal sensitivity. Whereas most existing methods often rely on dividing the time-series into larger segments of data (i.e. so called sliding window techniques), the point-based method (PBM) proposed here provides an estimate of brain connectivity at the level of individual sampled time-points. The achieved increase in temporal sensitivity, together with temporal graph network theory allowed us to study functional integration between, as well as within, resting-state networks. Our results show that functional integrations between two resting-state networks predominately occurs in bursts of activity with intermittent periods of less connectivity, whereas the functional connectivity within resting-state networks is characterized by a tonic/periodic connectivity pattern. Moreover, the point-based approach allowed us to estimate the persistency of brain connectivity, i.e. the duration of the intrinsic trace or memory of resting-state connectivity patterns. The described point-based method of dynamic resting-state functional connectivity allows for a detailed and expanded view on the temporal dynamics of resting-state connectivity that provides novel insights into how neuronal information processing is integrated in the human brain at the level of large-scale networks.

# INTRODUCTION

Studies of functional brain connectivity typically consider temporal correlations between different regions of the brain. Generally speaking, measures of connectivity are computed during a time window during which the participant is assumed to be at "rest". To this end, previous functional connectivity fMRI studies have shown that the connectivity profiles within the resting brain are segregated into several resting-state networks (RSNs) [Damoiseaux et al., 2006; De Luca et al., 2006; Fransson 2005; Fox et al., 2005; Greicius et al., 2003]. However, since resting-state fMRI brain activity is a manifestation of intrinsic, adaptive and dynamic processes that presumably, at least in part, reflect our continuous stream of conscious as well as subconscious, cognitive processes in our brains, there is no reason to assume that the degree of integration between individual fMRI RSNs is constant in time. Hence, rather than assuming resting-state fMRI data as being stationary in time, several recent studies have started to investigate the dynamics of resting-state fMRI activity (for recent review, see Hutchison et al., 2013a).

At present, the field of dynamic fMRI connectivity is still in its relative infancy where different approaches and methods have been proposed to isolate brain states with respect to their unique connectivity profiles. In the case of resting-state fMRI, three analytical strategies that can achieve this in the temporal domain have been proposed. First, temporal components (instead of spatial components) have been identified using independent component analysis [Smith et al., 2012]. Second, the dynamics of resting-state connectivity can be investigated using sliding window techniques [Kiviniemi et al., 2011; Hutchison et al., 2013b; Keilholz et al., 2013; Leonardi et al. 2013; Allen et al. 2014]. Third, several methods has been derived to target the non-stationary elements in BOLD signal intensity time-courses. For example, a study by Davis et al. 2011 examined the non-oscillatory aspects of resting-state signals by quantifying the relative variance of the amplitude of local-maxima and local-minima of signal intensity time-series. Using this variance-based approach, they could detect novel transient BOLD phenomena in sensory areas [Davis et al., 2011]. Similarly, transients of spatiotemporal dynamics of resting-state data were observed using a sliding correlation approach that detected recurrent, and rather brief periods of wide-spread patterns of correlative brain activity (Majeed et al., 2011). Furthermore, Tagaliazucchi et al. 2012 used a data reduction method based on a thresholding of the amplitude of the BOLD signal. Thus, by reducing the BOLD data to a spatiotemporal point process, they were able to show that resting-state networks can be derived from only a few data points [Tagliazucchi et al., 2012]. The idea that key aspects of the dynamics of resting-state networks can be adequately captured using only relatively small segments of the whole signal time-series was also targeted in a study by Liu and Duyn 2013. In that study, the authors showed that multiple resting-state networks could be reliably detected using, in some cases, as little as approximately 15 percent of the data. Additionally, it has been proposed that short spontaneous events contribute to functional connectivity [Allan et al., 2015]. Hence, results from previous studies suggest that the dynamics of resting-state fMRI signals may evolve at different time-scales, ranging from brief and transient events to oscillatory changes in brain activity. Thus, it would be advantageous to have a method that is sensitive to the proposed faster changes in dynamic functional brain connectivity.

The aim of the present work was to quantify dynamics aspects of neuronal integration between resting state networks. While all approaches to investigate the dynamics of brain connectivity have their own methodological advantages as well as disadvantages, we aimed to develop an analysis strategy that maximizes the temporal sensitivity since this would allow us to take advantage of the recent methodological developments within temporal graph theory [Holme and Saramäki 2012; Nicosia et al., 2013]. Importantly, temporal graph theory takes into account that connectivity (edges) between different regions of the brain (nodes/ROIs) may not always be present and may change over time. For example, while a person's entire social network could be represented in a static connectivity matrix, the temporal information of how frequent each person meets is lost. However, this information can easily be represented in temporal graph theory, where the existence of connections is also dependent on when two people meet. Temporal graph theory opens up the possibility to quantify the nature of network integration since it dynamically evolves in time. However, this requires a method is able to compute single time-point connectivity estimates as well as being sensitive to temporal fluctuations in brain connectivity.

In this paper we propose a new method to derive measures of dynamic functional connectivity and show how our proposed method can be used in analyzing the temporal properties of resting-state networks in the brain. Subsequently, we then explore two different dynamic properties of resting-state brain networks inspired by temporal graph theory. The first network property we explore is temporal burstiness in brain connectivity. Bursty properties in temporal graph theory have been applied to a wide range of social activities [Barabási 2005; Goh and Barabási 2006; Karsai et al., 2011; Takaguchi et al., 2013]. Theoretically, bursty processes are characterized by a higher occurrence of both shorter and longer durations of inter-contact times (i.e. the time duration between consecutive time-points of connectivity) than what is expected by chance alone (e.g. a Poisson distribution). Consider for example the simple case of a single dynamic, fluctuating degree of edge connectivity between two brain regions (nodes). Then, we say that the temporal pattern of connectivity between the two regions is bursty if the instances in time when the two regions are connected (inter-contact times) are characterized by a fat-tailed distribution where the probability of longer inter-contact times decays slowly. In contrast, a non-bursty process would either have a completely random presence of edge connectivity (i.e. follow a Poisson distribution), or possess a tonic/periodic connectivity pattern for which a presence of connectivity is either constant or oscillating over time with a small temporal variance between inter-contact times. The second network property we investigate in this study is persistency. Network persistency provides information regarding the temporal graphs by considering if there is a "memory" of the connectivity pattern within a network that extends over time. In other words, persistency tests how long there is, on average, still some similarity among successive temporal graphs. Together these two properties reveal different interesting temporal properties of resting-state brain connectivity and they illustrate how our proposed method for dynamic functional connectivity can be utilized to extract useful information regarding the dynamics of large-scale brain network connectivity.

**Materials and methods**

*Assumptions and motivation for the proposed method*

In the analysis of functional brain connectivity, we assume that the covariance between two regions of the brain reflects some degree of communication between them. In resting-state fMRI, we sample many time-points of data between two or more regions and then in a subsequent step compute the covariance between each region. In dynamic functional connectivity, we consider fluctuations in covariance between brain signals. One of the most popular methods to examine the dynamics of fMRI brain connectivity is the sliding window method. For the sliding-window approach, a window defines time-points that are next to each other in time to compute time-dependent estimates of signal covariance. One concern for the sliding window method in fMRI is that it requires windows that are quite long which raises the question: how temporally sensitive are we to the underlying dynamics of the brain if one minute of data is required to estimate it? Obviously, many dynamic changes in brain connectivity may occur during that time, even when taking into account the sluggishness of neuro-vascular BOLD response.

Given this background, we optimally want a method for dynamic functional connectivity analysis that provides us with an estimate of connectivity that has high temporal sensitivity. But, in order to achieve this, we need to sample multiple time-points to get a robust estimate of the signal covariance. These two requirements are difficult to reconcile in a single method, and in the case of the sliding-window approach it becomes a trade-off between the two, since increasing the temporal sensitivity (i.e. reducing the window length) will lead to a decrease in the amount of points used in estimating the covariance, which comes at the cost of more noisy estimates of the covariance.

In this paper we propose the foundation for a point-based method (PBM), a framework that aims to provide us with a high temporal sensitivity without the drawbacks that are inherent to the sliding-window approach. The logic used here is to remove the restraint that the estimates of covariance require time-points to be "neighbors" to each other in time. Instead we propose to gather time-points to estimate the covariance in a different manner: by identifying clusters that consist of time-points that have a similar global spatial space pattern (the spatial space defined here is spanned by the BOLD signal intensity time-courses from all ROIs used in the analysis) and estimating the covariance for each of these clusters. Thus, the PBM approach clusters time-points into similar spatial patterns (which we call "states"). Hence, the proposed strategy gives us many time-points in each state that can be used to estimate the co-variance. Each time-point in the resting-state fMRI session is thus assigned to the connectivity values that are representative for each cluster in the global spatial ROI space. Hence, the PBM approach provides dynamic connectivity estimates that are not necessarily based on neighboring sample points in time and thus results in a higher temporal sensitivity. An illustration of the proposed idea of assigning non-neighboring time-points to different clusters as outlined here is given in Supplementary Figure S1 which shows the procedure in the simple case of only two nodes/ROIs. But maximizing the temporal sensitivity as outlined above comes at a cost, namely that the PBM only yields a small number of possible connectivity estimates – a limitation that originates from the number of states chosen and that each state can only be associated with a specific value. But we would like to stress the fact that the limitation to only use a discrete set of connectivity values is not unsolvable since the proposed method can be extended to employ weighted connectivity matrices. This line of methodological development of the PBM method will be addressed in follow-up work and the limitation is also further addressed in the discussion section below.

*Data used, regions and subgraphs of interest, pre and post-processing.*

The 100 subject dataset from the Human Connectome 500 subject release was used (van Essen et al., 2012; Smith et al., 2013). We used the first resting state session with the RL (phase-encoding gradient in the right-left direction) encoding. Further information regarding the MR acquisition parameters can be found in Ugurbil et al. 2013 and van Essen et al., 2012. The downloaded data had undergone image preprocessing and removal of artifacts from non-neuronal origins by means of the FIX ICA (FMRIB's Independent Component Analysis-based X-noisifier) data artifact rejection process which removed ICA components from the data that were considered to constitute signal contributions from white matter, cerebro-spinal fluid, head movement, cardiac and respiratory sources [Glasser et al., 2013; Smith et al., 2013; Griffanti et al., 2014; Salimi-Khorshidi et al., 2014]. The data consisted of 1200 time-points per subject (TR = 0.72).

A parcellation scheme containing 264 spherical Region-of-Interests (ROI) with a radius of 10 mm defined along cortex and sub-cortical nuclei was used [Power et al., 2011]. Each ROI was assigned to a resting state network (or subgraphs) (we here employed the reduced 10 resting-state networks template; see Cole et al. 2013). For the purpose of completeness, the 10 resting-state network definitions used in the present investigation are given as follows: DM – default mode (58 nodes), SM – sensorimotor (35 nodes), Vis – visual (31 nodes), FP – fronto-parietal attention (24 nodes), Sa – saliency (18 nodes), CO – cingulo-opercular (14 nodes), Au – auditory (13 nodes), Sub – subcortical network (13 nodes), DA – dorsal attention (11 nodes) and VA - ventral attention network (9 nodes).  A scheme that contains their topographical layout onto the brain surface is provided in Supplementary Figure S2. For a more detailed picture of the spatial relationship between the location of the ROIs and brain anatomy, we refer the reader to Figure 3A in Cole et al., 2013.

To minimize the effect of movement [Power et al., 2012; Van Dijk et al., 2012], we performed image scrubbing by identifying time-points effected by movement using the framewise displacement (FD) method (rejection when FD>0.5). In brief, the FD method uses the information obtained from the six motion-related regressors created during image realignment. The FD value for each image volume is computed as the absolute value of the sum of the differences in motion between consecutive time-points. Further details can be found in Power et al., 2012. These data-points were deleted and the missing data time-points were estimated using a cubic spline interpolation. On average, 14.6 time-points (volumes) of a total of 1200 volumes per subject were marked as bad, removed, and interpolated. A band-pass filter (0.01-0.1Hz) was subsequently applied to the data, which is the frequency band in which the great majority of the resting-state brain activity resides. Each ROI's time-series was subsequently demeaned and transformed into Z-values by subtracting the global mean and dividing by the standard deviation. The time-series of data from all subjects was then concatenated along the temporal dimension.

*A point-based method to study dynamic functional connectivity.*

As a first step, we applied a Principal component analysis (PCA) to the fMRI ROI signal intensity time-series which reduced the spatial dimensionality of the data to 67 dimensions, which accounted for 85% of the variance in the data. This data reduction step was done with the intent to improve the accuracy of clustering since too many dimensions might hamper the clustering performance. K-mean clustering was performed to cluster each time-point along the reduced spatial dimensions for *k* values between 2 and 19. For each value of *k*, we performed up to 1000 iterations to reach convergence, and each choice of *k* was repeated 20 times to reduce any differences in clustering imposed by initial conditions. We opted for *k* = 8, a choice that resonates well with the choices made in previous work that have used clustering techniques to investigate dynamic fMRI connectivity [Allen et al., 2014; Liu and Duyn 2013]. Our justification for the choice of k was based on a local maximum of the silhouette plot (not shown) and further justified using a separate dataset for which similar spatial patterns were obtained (see below). To further minimize the possibility that our results are partly influenced by our choice of k, some of the key results were also reproduced using other choices of k (k=5 and 12, see Supplementary Figure S7).

*Creating s-graphlets and t-graphlets.*

We adopted the term graphlet to characterize our partitioning of the brain's dynamics by creating multiple connectivity matrices over a non-nodal dimension. The nodal dimension here is spatial (i.e. each ROI/node represents a different brain region). The term graphlet is a useful concept when analyzing fluctuations of graph metrics over some non-spatial variable (see Thompson & Fransson 2015a for an example of the usage of graphlet framework in the frequency domain). We derive graphlets that represent connectivity at different states and at different times. Along the motivation given above, we here use the term state to signify of an ensemble of time-points that share a similar global pattern along all the spatial dimensions defined by the previous clustering step. This entails that a particular state could be characterized in relational terms such as for example "a state for which the amplitude of nodes (ROIs) in the default mode network that are generally higher than the nodes (ROIs) in other networks" (examples of this relational description of states is provided in the Results section).

State-graphlets (s-graphlets) express the connectivity patterns in each of the k states.

They were constructed by taking the Pearson correlation coefficient between each of the 264x264 ROI pairs for time-points that belonged to each cluster. This procedure yielded 8 s-graphlets and as per convention with the chosen ROI template, the correlation between ROIs with centers less than 15 mm was set to 0, a step taken to prevent an abundance of local connectivity [Power et al., 2011]. Temporal-graphlets (t-graphlets) were subsequently constructed as a time-series of connectivity matrices for which each time-point was assigned to a s-graphlet, see also the illustration given in Figure 1. The assignment of a s-graphlet to a particular t-graphlet was dependent on which cluster that time-point belonged to in the clustering step. For each subject, 1200 t-graphlets were created, each with the size of 264x264 (representing the nodes). Each edge on a t-graphlet thus has k possible values attached to it, which can be viewed as a disadvantage of the proposed method. However, we again would like to emphasize the fact that a weighted connectivity matrix version of the point-based method can beyond a simple mapping of an s-graphlet to a t-graphlet and would thus provide a unique connectivity estimate for each time-point (see also discussion).

**Figure 1**

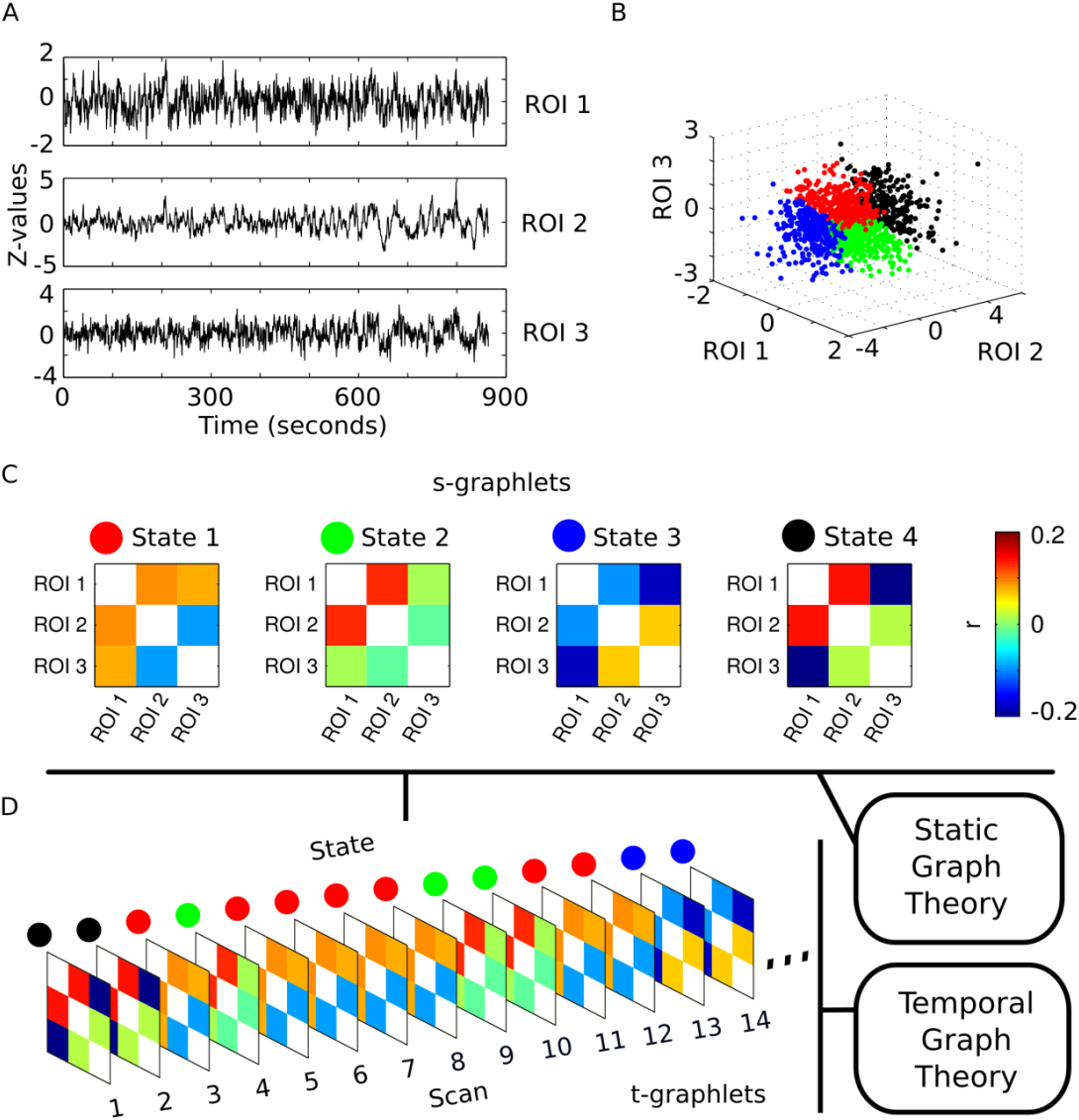

*Figure 1*. A schematic toy example that serves to illustrate the methodological framework used in this paper. (A) Z-values from three different regions of interest (ROI) during a single resting-state scan (demeaned and normalized by subtracting the mean and dividing by the standard deviation). (B) A 3D representation where each time-point is presented as a point in a three-dimensional space where each dimension represents the z-values of each of the three ROIs. Subsequently, k-mean clustering is performed which results in four clusters (color-coded). The individual clusters represent different

*states based on the underlying similarities in cluster space that is spanned by the time-series extracted from the three ROIs. (C) For each cluster derived in the step shown in B, a connectivity matrix between all ROIs is computed by means of a Pearson correlation that is based on all the time-points that belonged to that particular cluster. We name each connectivity matrix a state-graphlet (s-graphlet) on which, if desired, standard (static) graph theoretical measures can be computed. (D) In a last step, we create a time-series of s-graphlets where they act as representative connectivity templates at each instance in time. We call the time-series of connectivity matrices temporal graphlets (t-graphlets). Lastly, on the t-graphlets, we apply temporal graph theory to compute dynamical properties of functional resting-state connectivity.*

*Transition Probabilities.*

The probabilities of transition from one state to another were derived from each subject's time-series of connectivity matrices, i.e. their time-series of t-graphlets. Only time-points when a transition between states/s-graphlets occurred were taken into account. Hence, if two consecutive time points were assigned to the same s-graphlet, no transition was deemed to have taken place. The probability for each state/s-graphlet transitioning into each of the other seven possible s-graphlets was calculated.

*Sliding window analysis.*

For ease of reference, and to qualitatively compare the temporal sensitivity of the derived states, sliding window analysis was performed on the data for the purpose of comparing the state transitions found using the point based analysis versus the sliding window approach. For the sliding window analysis, the resting-state fMRI data had undergone identical post-processing steps as outlined above. Further, we calculated the covariance between each of the 264 ROIs for each subject for a window length of 120 time-points (86.4 seconds), comparable to the window lengths used in the previous sliding window literature [e.g. see Hutchison et al., 2013b; Allen et al., 2014]. Selecting the optimal window length is a balancing act between achieving a high SNR, i.e. long windows that entail high correlation values with a larger error versus shorter window lengths that provide good temporal accuracy but with low correlation values (i.e. low error) [Thompson et al., 2013]. Next, the time window was slid one time point down the time-series and a new connectivity matrix was computed for each step. To facilitate a comparison in terms of temporal sensitivity with the sliding window method, we used the k-mean clustering algorithm which was applied to each subject to classify connectivity matrices (again, we set $k = 8$). Since the dimensionality was 34716 x 1081 (number of edges x time-points) we chose to reduce the data dimensionality using PCA prior to the clustering of states. 30 spatial PCA components were retained, which accounted for more than 95% of the variance in the data.

*Matrix distance and persistent graph properties.*

The similarity between time-dependent connectivity matrices (t-graphlets) can be calculated in numerous ways. We chose to use the taxicab distance to measure the degree of similarity between t-graphlets. The taxicab distance, *d*, between two different t-graphlets *A* and *B* is defined as:

$$d_{AB} = \frac{1}{2}\sum_i^N \sum_j^N |A_{ij} - B_{ij}| \tag{1}$$

where *N* is the number of nodes in the network. The division by 2 is included in the expression for case of the undirected graphs. The smallest possible *d*, if both A and B are identical, is 0. The highest possible value of *d*, for an undirected matrix with values ranging between -1 and 1 with self-edges set to 0 is $(N^2-N)/2$. To facilitate a possible comparison between t-graphlets of different sizes, persistency was scaled between 0 and the maximum possible distance. For other methods of computing matrix temporal similarity, see Nicosia et al. 2013. Apart from computing persistency, the normalized taxicab distance was also used for comparing similarity between s-graphlets.

We define persistency as the average length of time when some degree of similarity (i.e. when the distance is small) exists between the t-graphlets A and B, where B is positioned Δt time-points away from A. We computed the degree of persistency for a range of temporal distances, i.e. Δt ranged between 1 and 50 TRs (TR = 0.72 seconds). The distances were averaged across nodes and subjects for every value of Δt. The persistency of connectivity time-series of t-graphlets is then considered to be the average distance that is smaller than the statistical significance threshold (see below) and thus still, on average, shows significant similarity with the connectivity profile Δt time-points before. This provides an estimate of the average duration of time after a given time-point when there is still some degree of similarity in connectivity that is larger than chance alone.

*Computing Burstiness.*

Burstiness measures whether the presence of network edges, in our case occurrences of brain connectivity in time, are characterized by an occurrence of very brief inter-contact times intertwined with long and varying inter-contact times. For a given edge in a network, the inter-contact period is defined as the time-period between consecutive presences of an edge in the network. To test whether the dynamics of brain connectivity possess bursty temporal characteristics, we first thresholded the t-graphlets so that only edges that displayed a strong degree of connectivity were considered in the analysis. Some kind of thresholding of edges is required in order to determine which edges are "on" and "off" and here we chose to threshold proportionally (i.e. keeping a certain percentage of all edges) in each t-graphlet. To demonstrate the robustness of our method with respect to the chosen multiple thresholds, we calculated edge burstiness by retaining the top 5 % and 10 % of all edges, respectively. Next, the thresholded t-graphlets were transformed to binary matrices where a 1 represent a presence of a given edge and 0 an absence of the edge at a given time-point. The time-series of binary t-graphlets is then the basis for computing inter-contact periods of all edges. As a final step, the inter-contact times were pooled over subjects. The measure of burstiness used here was taken from Goh & Barabási [Goh and Barabási 2006; Holme and Saramäki 2012]. The burstiness of an edge between two nodes is calculated by

$$B = \frac{\sigma_\tau - \mu_\tau}{\sigma_\tau + \mu_\tau} \qquad (2)$$

where $\sigma_\tau$, and $\mu_\tau$ are the standard deviation and the mean of the inter-contact times, respectively. Burstiness is calculated for each edge in all subjects and its value ranges between -1 and 1. A value that is greater than 0 suggests a bursty pattern of brain connectivity. Values less than zero imply a more tonic/periodic pattern of brain connectivity with less variability in inter-contact times. Values centered at 0 suggest that connectivity occurs randomly over time. Permutation tests were carried out (see below) to determine whether edges were significantly bursty or tonic/periodic.

*Statistical testing.*

All statistical comparisons were derived by Monte Carlo non-parametric tests where the empirical results were compared to distributions created by shuffling one of the variables of the empirical data. The shuffling procedure outlined here followed one of the standard practices for creating null models in temporal graph theory (Holme & Saramäki 2012). The procedure to shuffle the data to create null distributions for graph theoretical parameters was slightly modified with regard to the parameter being tested and below we go through each shuffling procedure in turn.

To test for significant differences in connectivity among s-graphlets, the null hypothesis for this test was that there was no difference in the connectivity between the states derived through the k-means clustering. Each comparison between two clusters entailed a computation of 10,000 permutations in which the time-points between two clusters (states) were randomly assigned to two random groups, preserving the number of time-points that matched the original clusters. At each permutation, s-graphlets were derived for the two clusters being compared. A distribution of permuted differences was created by subtracting one clusters' permuted s-graphlets from the other. Each edge was then compared to the difference between the its distribution from the distribution of permuted s-graphlets ($p<0.001$, two tailed, FDR-adjusted for all edges and s-graphlet comparisons).

For the statistical testing of persistency, the null hypothesis for this test was that there was no similarity between t-graphlets situated Δt apart. To create the non-parametric distribution, the entire t-graphlet time-series, for each subject, had their order of the time-points randomly shuffled. This was done for 200 permutations. The same procedure of computing the taxicab distance between t-graphlets were carried out for all selections of Δt. Persistency was determined as the highest Δt that had a smaller distance than the permuted value ($p<0.05$, one-tailed).

To test for significant burstiness of brain connectivity, the null hypothesis was that the distribution of inter-contact times followed a Poisson distribution that contained neither tonic/periodic nor bursty inter-contact times. The non-parametric test was done by creating a distribution using 200 permutations where the order of the binary time-series (thresholded t-graphlets) was shuffled to create new inter-contact times for each permutation. This was done for all possible edges. For the purpose of creating distributions to which all edges could be compared and to circumnavigate the multiple comparison problem we, at each permutation, used the maximum and minimum permuted value of the bursty coefficients, taken over the all edges, to create the distributions with which to compare the empirical values to. For an edge to be considered to be bursty, its value had to be above 0 and equal or higher than the 195[th] highest permuted value from the maximum distribution ($p≤0.05$, two-tailed). For an edge to be considered periodic, it had to have a value of less than 0 and equal or lower than the 5[th] lowest permuted value from the minimum distribution ($p≤0.05$, two-tailed). We used a significance threshold of *p* less or equal than 0.05 here, due to the possibility of all permuted values and the empirical values being -1 (i.e. the edge connection is always present). We limited the number of permutations to 200 for both burstiness and persistency permutations tests due to computational restraints.

Finally, we tested whether there was a difference in the average network bursty coefficient for between-RSN versus within-RSN edges. Here, the null hypothesis was that there was no difference in the average between-RSN and within-RSN bursty coefficients. To test this, a non-parametric permutation test was created by shuffling whether a network edge was considered to be within or between- RSNs. To test whether the average bursty coefficient for between-RSN connectivity was significantly different to the average within-RSN coefficient, randomly assigned each edge to two groups, preserving the number of between-RSN edges in the first group and the number of within-RSN edges in the second group. The empirical difference between average between-RSN and within-RSN values was then compared to a distribution of the differences of the averaged values in 1000 permuted groups. The difference was classified as significant if it was larger than the $975^{th}$ permutation ($p<0.05$, two tailed).

*Justifying the number of chosen clusters (k) using an independent dataset.*

It first needs to be said that in any type of clustering analysis performed on fMRI data when the number of clusters (*k*) has to be empirically determined, there are two problems which might occur. (1) The value of k is set too low, resulting in a too small number of clusters being defined which would potentially miss out on some aspects of the brain dynamics. (2) The value of k is set too high which results in too many clusters being defined, leading to an "over-splitting" of the fMRI data. The case of over-splitting could prove to be highly detrimental, as an additional split of a cluster could possible result in that the magnitude of correlation between the BOLD signal intensity values in a cluster switch from being positive to negative (or vice versa), inducing spurious connectivity. Thus, providing as solid experimental support as possible for the chosen value of k is of paramount interest. As an alternate route, the analysis can be extended to include a validation of the results for a range of values of k to show a robustness of an effect that is largely independent on the size of k.

In this work we have both developed a strategy that includes a separate, independent data to validate our choice of k as well as validating our results for different choices of k. With respect to the first approach, we performed additional analyses on the second dataset from the human connectome dataset (i.e. the so called "LR" dataset from the first resting scan) obtained from the same subjects. Identical pre-processing and post-processing steps were applied. A schematic diagram of the work-flow used here is presented in Supplementary figure 3A. Prior to clustering, the complete dataset was evenly split into three blocks of data. The time-points (or equivalently image volumes) 1 to 400 for all subjects were placed in the first block (here named A), 401 to 800 were placed in the second block (B), and time-points 801 to 1200 was allocated to the third block (C).

For all three data blocks, a clustering using k-means algorithm was computed for $k$ =2 to 20. In the next step, for each $k$, we paired the clusters between each combination of block dataset (i.e. possible parings in this case are AB, AC and BC) based on the average BOLD activity for each node in each cluster (see Supplementary Figure 3B that shows an example for k = 8). The paring was achieved by computing distance matrices between each data block's clusters using the averaged taxicab distance (see equation (1)) for the average BOLD activity of each node. This procedure yielded a $k$ x $k$ distance matrix for each unique paring of block dataset. Subsequently, the so called Hungarian Algorithm [Kuhn 1955; Munkres 1957] was used to find the optimal cluster pairing for each distance matrix. The Hungarian Algorithm was applied to the clustering paring distance matrices and the output is a $k$ x 2 solution matrix for each of the block dataset pairings. Since there are three possible block dataset pairings, it is conceivable that the optimal solutions for the different block dataset pairings are not transitive and thereby resulting in incompatible clustering solutions. This possibility was accounted for by checking if the pairing of clusters in the block dataset pairing AC was identical to the pairing of clusters that could be derived for AC by using AB and BC. After checking for transitivity among possible pairing of data blocks, nine of the tested k values had transitive solutions (k = 2, 3, 4, 5, 6, 7, 8, 11 and 12). This suggests that any choice of k that has a transitive solution is to some extent reproducible in terms of k-means clustering with similar spatial patterns independently in different parts of the second data set. Hence, by the validation of our approach using an independent data set, we believe that our choice of setting k=8 is justified, since it presented us with similar global spatial patterns for the three different parts of an independent dataset. Additionally, our key results were replicated using other choices of k (for results setting k to 5 and 12, see results shown in Supplementary Figure S7). It should be noted that our justification of choice of k that includes the splitting of an independent dataset into three parts relies on the assumption that the number of derived states, in the same subjects will identical.

# RESULTS

*Outline of the point-based method of dynamic functional connectivity in a simple dataset.*

To illustrate our method, we exemplify the analysis by the use of a small, toy data set shown in Figure 1 (for all full list of all pre-processing steps taken, see *Methods and Materials*). In this simple example, we use the fMRI BOLD time-series extracted from three regions of interest (ROIs) in a single subject (Figure 1A). All three dimensions in the state-space of the three ROIs under consideration are treated independently and all signal time-points in the time-series are used for the cluster identification procedure that aims to identify clusters based on their co-activation pattern in all three dimensions. Although any clustering algorithm can be used, we have here chosen the k-means clustering method for computational reasons and $k$ was in this toy example set to $k = 4$. Different time-points can be grouped together based on similar global co-activation profiles of the spatial dimensions, regardless of their temporal location (Figure 1*B*). We name the clusters that have a similar spatial organization for a state. It is important to emphasize the fact that the clusters obtained from the clustering of the full data set that are described below (264 spatial dimensions, reduced to 67 spatial dimensions after PCA), do not align themselves into easily recognized groups as shown for the toy example in Figure 1B, but span complex trajectories in the high-dimensional spatial space.

For each state, a connectivity matrix is created that is based on all the data time-points belonging to that cluster (Figure 1*C*). We call the corresponding connectivity matrix, for a given state, a state-graphlet or *s-graphlet.* Thus, a s-graphlet can be viewed as a connectivity matrix that represents the existing global co-activation pattern computed on all time-points that belong to the given cluster. Given the assumptions of functional connectivity, each correlation between two ROIs in a s-graphlet represents the connectivity when a given "state" (i.e. the global activation pattern of all ROIs) is present. It can be seen in Figure 1C that the connectivity profiles among the four s-graphlets are quite different from each other. For example, the connectivity values between ROIs that are positive in s-graphlet 1 turn negative in s-graphlet 3 and vice versa. It should be noted that all the usual graph-network theory can be applied on s-graphlets (e.g. degree centrality, betweenness centrality and community detection, see Bullmore and Sporns 2009). Finally, each time-point is assigned to the connectivity matrix of the state/cluster to which it belongs (see also Supplementary Figure S1). Thus, every time-point is represented as a single s-graphlet in a time-series of connectivity matrices (Figure 1D). By constructing a time-series of connectivity in this way, methods from temporal network theory can be applied to compute temporal measures of connectivity, e.g. graph burstiness and persistency. Thus, to emphasize the fact that the sampled time-points are now represented in the form of individual s-graphlets, we have chosen to use the term t-graphlets (temporal-graphlet) for the corresponding s-graphlets when they are inserted and put together to form a time-series of connectivity matrices. This means that measures of the dynamics of resting-state networks, such as network burstiness and persistency are computed on the temporal sequences of t-graphlets. The construction and configuration of s- and t-graphlets are shown in Figures 1C and 1D.

*Construction of s-graphlets.*

We now leave our illustrative example and turn our attention to the full data set. As mentioned previously, it is important to note that the clustering of signal intensity time-courses resulted in states that spanned across subjects and hence did not end up in a trivial partitioning, e.g. all time-points (image volumes) from one subject assigned to one state and another subject into a second state and so forth. Instead, the clustering resulted in a distribution of different states on the subject-level across all time-points (Figure 2*A*). The distribution of states differed to some extent between subjects, which is to be expected, but all states were represented at the group level as shown in Figure 2B. Moreover, Figure 2*B* shows that while some states occurred relatively more often (state 6, 7 and 8) others occurred relatively more seldom (states 2 and 5).

**Figure 2**

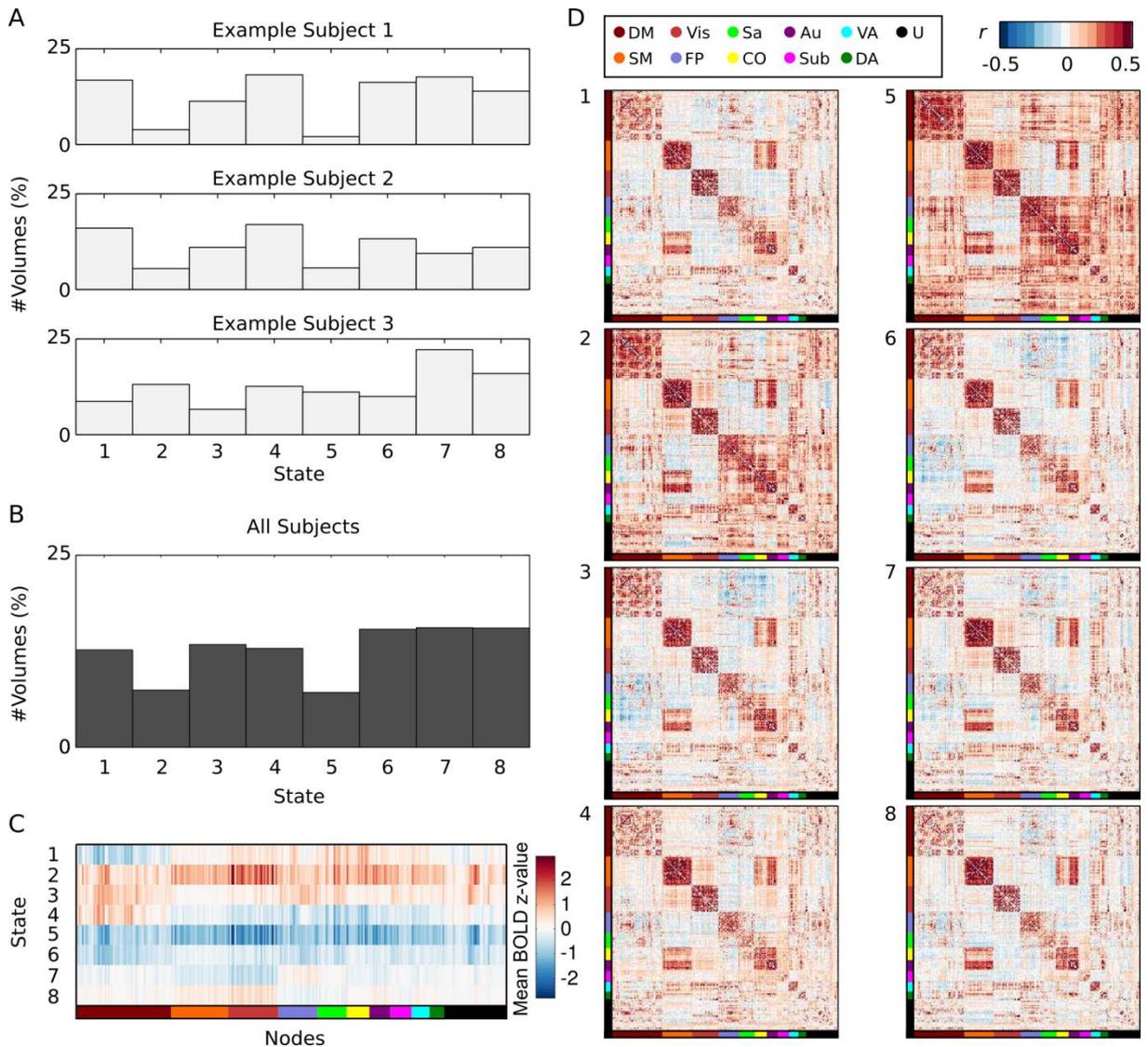

*Figure 2. Distribution of image volumes (or equivalently, time-points) across states (s-graphlets) and subjects. (A) Percentage of image volumes in three representative subjects that were assigned to each of the eight states by means of k-mean clustering (number of clusters, k = 8). (B) Percentage of image volumes for all subjects that were assigned to each of the eight states. (C) Average Z-value over all assigned time-points in each state for each node. The colored bar at the bottom of the figure corresponds to the nodes assigned to each network. (D) Connectivity matrices for all states/s-graphlets using time-points concatenated across all subjects. The nodes in each connectivity matrix are ordered according to their designated resting-state network (DM – default mode, Vis – visual, Sa – Saliency, Au – Auditory, VA – ventral attention, SM – sensorimotor, FP – fronto-parietal attention, CO – cingulo-opercular, Sub – Subcortical, DA – dorsal attention, U – unassigned).*

Since the states (s-graphlets) were derived from a clustering the BOLD signal intensity time-series of the ROIs (reduced to spatial PCA dimensions in preprocessing step), it is of interest to consider how the amplitude of the BOLD signal varies between states. The average BOLD z-values for each ROI are shown in Figure 2C for each of the eight derived states/s-graphlets. From Figure 2C, some global trends can be observed. For example, in the slightly less occurring states (2 and 5, see Figures 2A-B), all the ROIs have their lowest and highest amplitudes respectively. Moreover, for state 4, the nodes/ROIs residing in the default mode network have on average higher BOLD amplitude than all other nodes. Further, state 1 shows the opposite trend with on average lower amplitudes for nodes within the default mode network than all other nodes. However, great caution in interpreting the amplitudes shown in this panel is warranted. We need to remind ourselves that the k-mean clustering algorithm operates on clustered time-points based on the PCA derived spatial dimensions. It is important to point out that this procedure does not result in that the BOLD amplitude of each ROI time-series is partitioned according to its signal intensity since each individual time-point is assigned to clusters based on the global state across all spatial dimensions simultaneously. Rather, the information given in Figure 2C should be interpreted as the pattern of the ROI BOLD signal intensities in relation to each other. This interpretation is justified when considering the variance of the amplitude of a ROI signal intensity time-series. As an example, by just glancing at Figure 2C, one get the impression that the maximal amplitude for a given time-series is always found in state 2 and the minimal amplitude is found in state 5. That this is not the case is exemplified in Figure 3 that shows the state/s-graphlet membership for each time-point in three different ROIs in a representative subject. Two important properties of the point-based method to estimate dynamic functional connectivity are highlighted in Figure 3. First, the minimal and maximal amplitude values for different ROI do not automatically belong to the same state. Second, state memberships of time-points are distributed along the time domain in a non-trivial way. This means that at a given instance in time, for example around TR = 240, the dynamics of the resting-state brain is assigned to state 3 (light green in Figure 3). At the same time, the BOLD amplitude is in a local minima for two ROIs (middle and lower panel in Figure 3) and at local maxima for one ROI (upper panel in Figure 3). Another example is at approximately TR = 380 when the brain dynamics is assigned to state 2 (orange in Figure 3) which for ROI 1 and ROI 3 imply a local maxima in BOLD amplitude whereas for ROI 2 it is in a local minima. Thus, the behavior shown in Figure 3 suggests that the k-means clustering is based on the global relation between the ROI signal intensity time-courses rather than the intensity for each individual ROI per se.

**Figure 3**

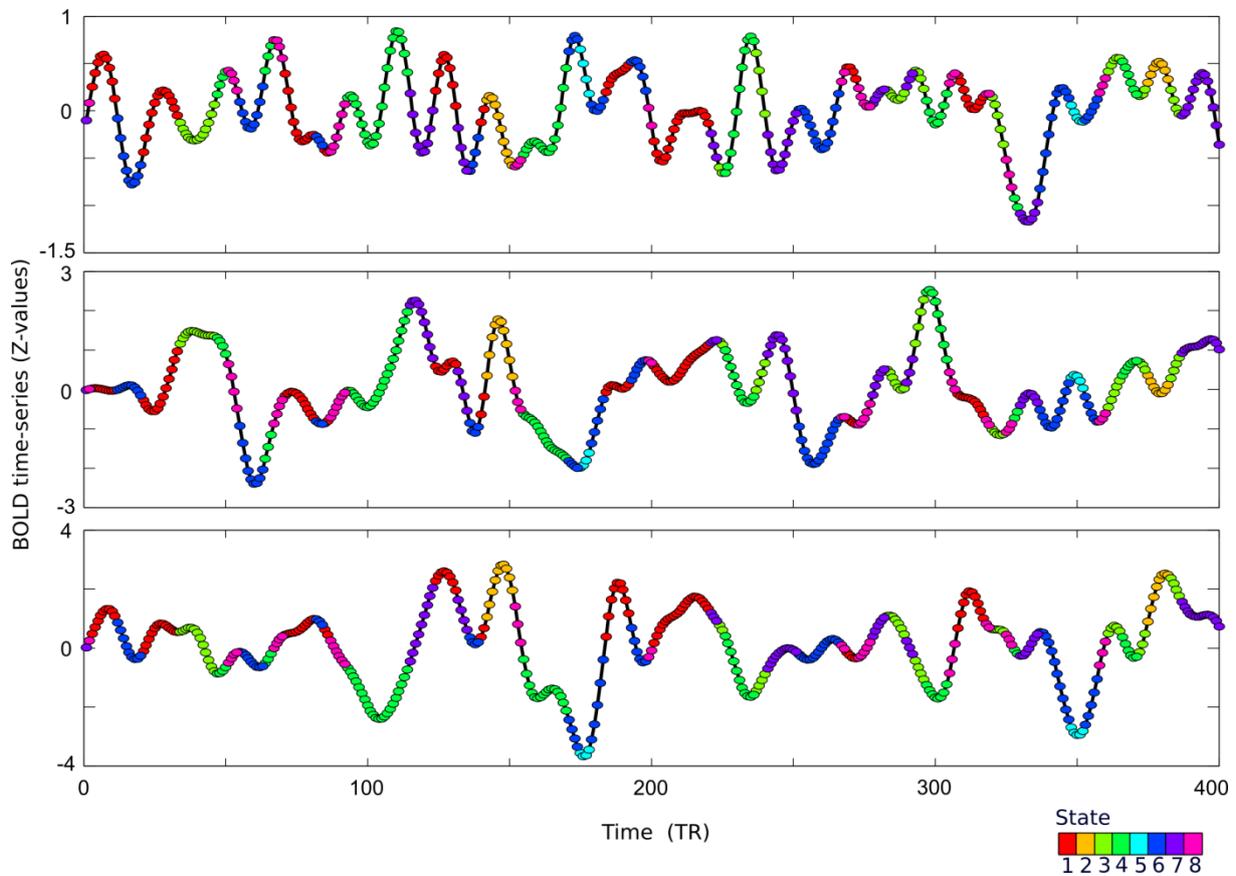

*Figure 3. Assignment of states for resting-state fMRI data where state (color-coded) is plotted on top of the standardized BOLD time-series for three different ROIs in one representative subject (only the first 400 time-points are shown). Each BOLD signal intensity time-point is assigned to one of the eight possible states by the point-based method. Here, it is important to point out the point-based clustering method does not result in a trivial partitioning of the time-series based on the amplitude of the BOLD signal. For an example, the first peak (t = 10 TR) in the first ROI (upper panel) with a Z-value greater than 0.5 is assigned to state 1 whereas a trough at t = 204 TR (Z-value less than -0.5) is also assigned to state 1. Many other examples of this behavior can be found in Figure 3, for example the second ROI (middle panel) has local trough at t=240 TR and a peak at t = 300 TR that are assigned to the same state (state 3). Hence, the point-based method yields a complex pattern of states that are based on the global features of the space spanned by the ROI signal intensity time-courses using the PCA components as input.*

For each state/s-graphlet, all time-points that belonged to the corresponding cluster were used to compute the connectivity matrices between all ROIs as shown in Figure 2D. Interestingly, the two s-graphlets with the least amount of time-points/image volumes attached to them (s-graphlets 2 and 5) show very visible increases in overall brain connectivity. Other differences between s-graphlets in Figure 2D are more subtle, but some differences can be observed, both near the diagonal, indicating increased within-RSN connectivity as well as more distant from the diagonal that indicates an increased degree of between-RSN connectivity. The differences and similarities in strength of connectivity within- as well as between-RSNs for the eight s-graphlets/states are further highlighted in the Kamada-Kawai spring-embedding visualization shown in Supplementary Figure S4.

One might consider the possibility that at least one of the states/s-graphlets correlates with movement, as micro head-movement is known to be a major concern for resting state connectivity studies [Power et al., 2012; Van Dijk et al., 2012]. The relative amount of motion-afflicted time-points for each state/s-graphlet is displayed in Supplementary Figure S5. Although motion-afflicted time-points are present in all states, it can be seen that state 8, and to some small extent also states 2 and 7 (see Supplementary Figure S5), had a slightly higher disposition for micro head-movement in relative measures compared to the other five states. So the question is if the relatively higher propensity for head-movement observed for state 8 should be taken as an indication it is a state that can be attributed to micro head-movements? When considering this possibility it is important to keep in mind that the numbers of motion-afflicted time-points provided in Supplementary Figure S5A are given as a percentage of the total number of time-points across subjects that had a FD-value higher than 0.5. Importantly, the average (across subjects) number of time-points for which the FD-value was larger than the chosen threshold was 14.6 time-points per imaging session (1.2 percent of the total of 1200 time-points). For example, this implies that the number of approximately 330 image volumes deemed to be affected by micro-motion in state 8 as shown in Figure S4B corresponds, on average, to 4.32 time-points afflicted with micro head-motion per session (out of a total of 1200). So, although the amount of motion-afflicted time-points differs between states, the difference in absolute terms is quite small, which suggests that no particular state can be singled out as a state that is driven by micro head-movements.

*Comparison between s-graphlets.*

The question whether the derived connectivity matrices for the different states actually show significant differences in functional connectivity is of importance. As noted in our simple example shown in Figure 1, there is no built-in mechanism in the k-means clustering method that enforces s-graphlets/state connectivity matrices to be significantly different from each other. Hypothetically, the variance along the spatial dimensions for each state could be very similar between two clusters, which would result in s-graphlets that do not significantly differ from each other. To ensure that this was not the case, we tested for significant differences between all the s-graphlets using a data shuffling method (p<0.001, two-tailed, FDR-adjusted, see *Methods*).

Since the focus of our study was to examine the dynamics of network integration, we have chosen to present how the s-graphlets significantly differed from each other at the level of resting-state networks. A comparison of the degree of strength of connectivity for all pair-wise combinations of resting-state networks for all eight s-graphlets is shown in Figure 4. Each matrix shows the number of edges (expressed in percentage) that is significantly (permutation tests across edges and RSNs, see *Methods and Materials*) larger in one s-graphlet compared to another s-graphlet. For example, the color-coded matrix elements that together constitute the matrix depicted in the first column and second row in Figure 4 shows the number of edges between resting-state networks (in percentage, range: 0 – 100 %) that was significantly larger in s-graphlet 2 compared to s-graphlet 1. Correspondingly, the complementary matrix located in the second column and first row in Figure 4 shows the reverse contrast, namely the number of significantly stronger edges between and within resting-state networks in s-graphlet 1 compared to s-graphlet 2. Besides the overall impression that connectivity between-RSNs are considerably stronger in s-graphlet 2 compared to s-graphlet 1, we can from Figure 4 (see also the network description given in the lower right corner of Figure 4) observe differences at the level of individual resting-state networks. In our example, we can for example conclude that the degree of within-RSN connectivity for the default mode (DM) is significantly larger in s-graphlet 1 than s-graphlet 2 but at the same time is the degree of connectivity between the somatomotor (SM) and dorsal attention (DA) networks significantly larger in s-graphlet 2 compared to s-graphlet 1. Taken together, the results shown in Figures 3 and 4 strongly suggests that the k-mean clustering of the resting-state fMRI data into 8 states/s-graphlets have yielded a partitioning that separates the data into states that each have their own unique characteristics in terms of connectivity profiles, both at the level of individual edges between ROIs as well as for the ten a priori defined resting-state networks.

**Figure 4**

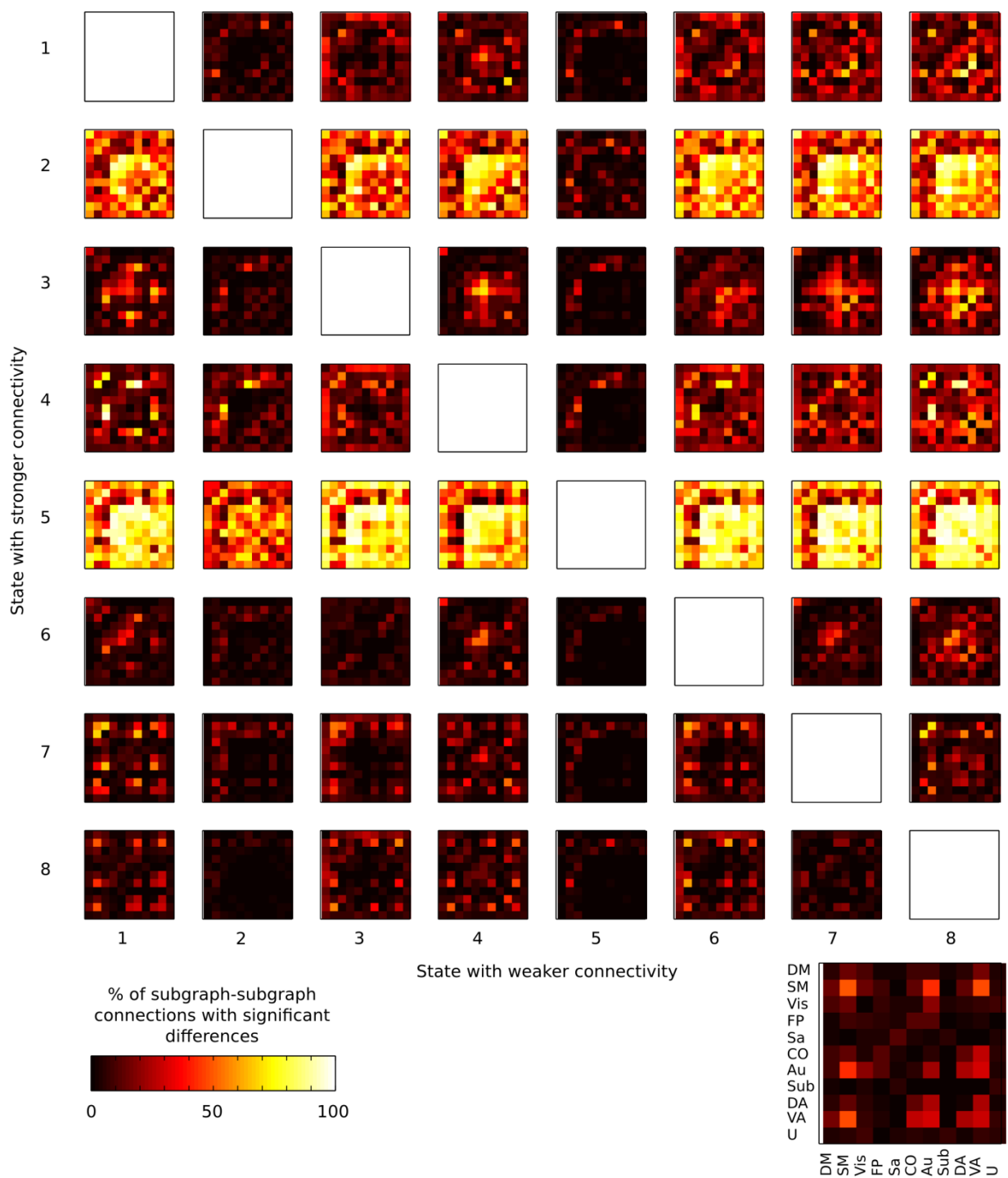

*Figure 4. Difference in connectivity between s-graphlets at the level of resting-state networks. The results show that all s-graphlets have significant differences between resting state networks. Percent of edges between different resting state networks for each s-graphlet comparison where an edge was significantly larger in one state compared*

*to the other (p<0.001) based on 10,000 permutations and all the p-values were FDR adjusted based on the p-values obtained for all edge s-graphlet comparisons.*

*The construction of t-graphlets, transitions between states and the probability of transition.*

After the derivation of the s-graphlets, we inserted each connectivity matrix into a time-series, such that each time-point takes the s-graphlet from the cluster wherein that time-point was classified. This procedure results in a time-series of t-graphlets (see also Figure 1*D*) for which each time-point is represented as a t-graphlet, which in turn is one of the 8 previously derived s-graphlets. Thus, we can describe the dynamics in brain network connectivity as a temporal arrangement of t-graphlets where the total number of included t-graphlets matches the number of sampled data points, in our case 1200 t-graphlets per subject/session. The achieved temporal sensitivity using the PBM approach is shown for three representative subjects in Figure 5A. For comparison, we performed a sliding window dynamical connectivity analysis (window length = 86.4 seconds) on the same data using an identical number of states (Figure 5B). Importantly, the comparison with the sliding window analysis was done with the intent to illustrate the different temporal scales of the two methods. Hence, we do not make any attempts to quantify the difference of state duration yielded by the two methods. Note that although the color scheme used to describe the procession of state membership as a function of time in Figures 5A and 5B is the same for both methods, this should not be interpreted that the actual states of functional connectivity are identical across methods. The procession of t-graphlets for the PBM and sliding window methods that are outlined in Figures 5A-B shows that far fewer state transitions are detected for the sliding window method and subjects tend to stay longer in each state compared to the PBM method. Moreover, the transition plots shown in Figures 5B are quite similar to what was derived in previous investigations using sliding window analysis (see for example Allen et al. 2014). In comparison, it is clear that the point-based method presented here is able to detect much faster state transitions with a substantially increased temporal sensitivity. As shown in Figure 5C, the average duration for staying in a given state/s-graphlet is 6.3 scans (4.5 seconds) with the point based method. Histograms that shows the full distribution of the duration for staying in a given state is shown for all eight states in Supplementary Figure S6.

In a subsequent step, we computed the state/s-graphlet transition probability for the point-based method, which essentially tells us the probability of transitioning from a given state/s-graphlet at time-point *t* to any other state/s-graphlet at time-point *t+1*.

The state transition probability matrix shown in Figure 5D reveals that some states/s-graphlets are more reachable than others. It is noteworthy that the two states/s-graphlets that show an apparent high degree of connectivity across RSNs (s-graphlets 2 and 5, see Figure 2*C*) are both highly likely to transfer into states/s-graphlets 3 and 6 respectively, for which the global pattern of strength of connectivity for the sensory and motor networks (e.g. SM, Vis and AU) are stronger in states/s-graphlets 2 and 5 whereas the frontal-parietal and default mode networks harbour the opposite pattern. Given that there is some overlap in many connections and the short time spent in each state/s-graphlet, we asked whether the probability of transitioning from one s-graphlet to another correlated with the similarity between states/s-graphlets. Using the normalized taxicab distance, we found that there is indeed a negative correlation between the probability of transitioning from one state/s-graphlet to another and the corresponding similarity in terms of distance between the two states/s-graphlets (Figure 5E, *ρ*=-0.3338, p=0.0119). This result speaks against the possibility that the duration between a change of state/s-graphlet is too short by the methodology used, since the probability for a state transfer to occur is larger when the distance between the two states/s-graphlets is less.

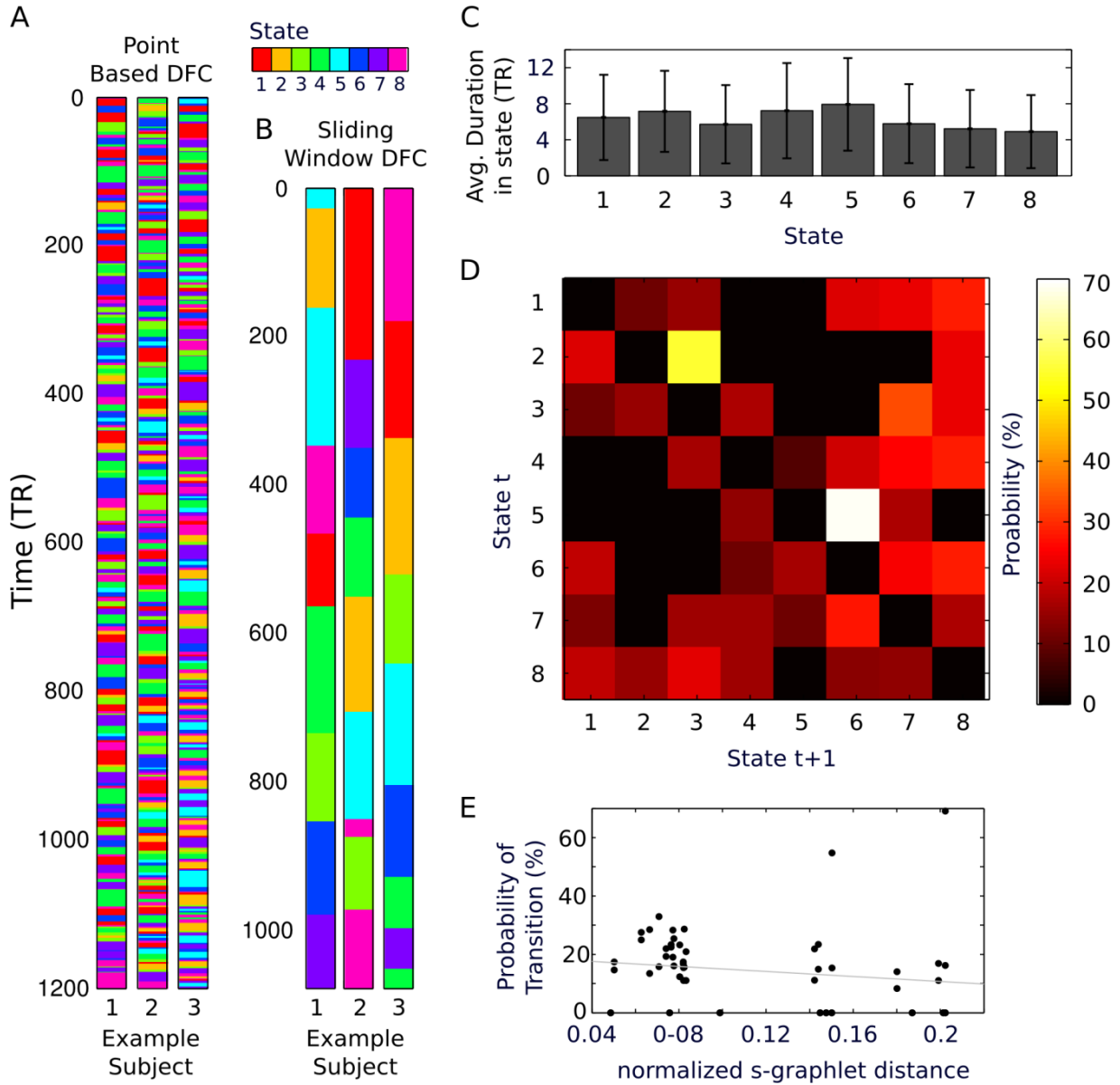

*Figure 5.* Transition of state in resting-state fMRI time-series. (A) Example of the temporal transition of s-graphlet/state in three representative subjects (time is here expressed in TR (scans) = 0.72 seconds) using the point-based method of dynamic functional connectivity. (B) Corresponding temporal transition of state in the same subjects using the sliding window method (window length = 86.4 seconds). Note that although the coloring scheme used is the same for the two methods shown in panels A and B, this does not imply that the actual state connectivity patterns are identical. (C) The average (across subjects) duration (again expressed in TR) a subject stays in any given state (1-8) before a transition of state occur. Error bars depict the standard

*deviation. (D) Transition probability graph showing the probability that a transition of state occur between time t and t+1. Each matrix row adds up to 100 percent. (E) The probability of the occurrence of a state transition versus the distance between two s-graphlets. The figure shows a significant negative correlation between the two (ρ = -0.3338, p=0.0119).*

*The persistency of brain connectivity.*

The high temporal sensitivity provided by the point-based method granted us an opportunity to quantify the dynamical properties of between- as well as within-network connectivity using methods developed within temporal graph theory. First, we asked whether we could find a time-limit for which there was still some significant resemblance to a previous temporal connectivity matrix, or in other words the persistency of edges across t-graphlets. Given the results shown in Figure 5*E*, we know that there is a general tendency of an increased probability to change state when the corresponding correlation matrices are similar. We were now interested to quantify the average degree of similarity over time for t-graphlets. For each subject and for every time-point, the normalized taxicab distance was calculated between each correlation matrix (t-graphlet) at time-point, $t$, and $\Delta t$ ($\Delta t$, ranging from 1 to 50 TRs). The taxicab distances were subsequently averaged over time-points and subjects and the results are shown in Figure 6. The plot shown in Figure 6 suggests that for any given time-point, the average time duration before the connectivity profile is no longer more similar than by chance alone is approximately 13.7 seconds (19 scans) (p<0.05). This observation is interesting for two reasons. First, non-time locked data, i.e. resting-state, shows a significant measure of persistency of co-activation that resembles that of the BOLD signal time-course obtained following an event or stimuli given in a typical event-related fMRI design. Second, at a given point in time, our persistency estimate provides an indication of how long the "memory" in terms of intrinsic brain connectivity patterns of the BOLD signal exists before it becomes longer in duration than the average length of any state and hence untraceable.

**Figure 6**

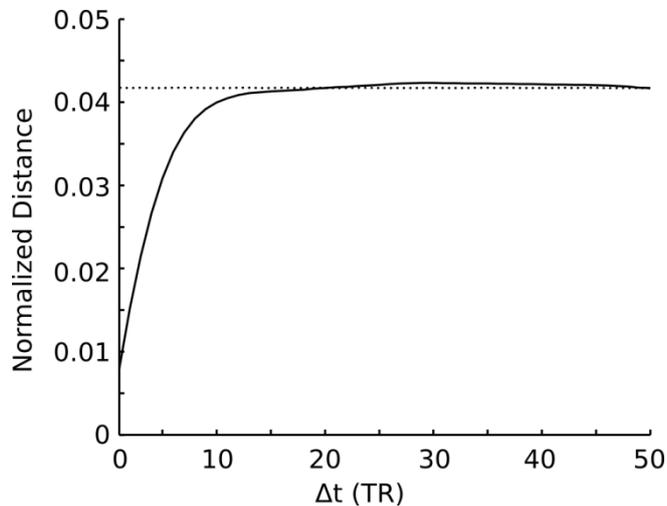

*Figure 6. The persistence of resting-state brain connectivity. The average normalized distance in terms of connectivity difference between two t-graphlets situated Δt scan apart from each other. Each t-graphlet is compared with every other t-graphlet positioned Δt time-points away, and the average normalized distance is computed. The dashed line shows the significance level based on computing the same metric with randomized time-series (p<0.05). The degree of similarity of connectivity strength between time-points that is larger than chance alone between individual t-graphlets is approximately Δt = 19 (13.7 seconds).*

*The burstiness of brain network integration.*

Next, we asked whether the point-based method could be useful to provide insight into which extent the dynamics of resting-state brain connectivity accommodate bursty modes of communication. Examples of bursty and periodic lists of inter-contact times from a single subject are given in Figure 7A. Another way to illustrate the bursty characteristics of brain connectivity is to compare the distribution of edge inter-contact times to a Poisson distribution. Figure 7B shows an example of the empirical distribution of inter-contact times obtained from a single edge (pooled over subjects) compared to a corresponding Poisson and an approximated Pareto distribution. By comparison to the Poisson distribution, the empirical distribution of inter-contact times shows a fat-tail at longer connect times together with large amount of brief inter-contact times. Together, these deviances from a Poisson distribution are hallmark features of a temporal process that has a bursty characteristic. Using the measure to quantify burstiness given in [Goh and Barabási 2006; Holme and Saramäki 2012; see also Methods and Materials], we calculated the percentage of edges that showed either significantly bursty or periodic patterns (p≤0.05, two-tailed). At both edge thresholds used (top 5 and 10 percent respectively), the results shown in Figures 7C-F suggest that within-RSN connectivity (diagonal elements) contains a mixture of both periodic and bursty dynamic connectivity patterns. Perhaps rather unsurprisingly, a substantial part of within-RSN temporal connectivity, in particular so for the lower 10 percent edge threshold, is characterized by periodic patterns of connectivity. Although the between-RSN connectivity profiles also display a mixture of both tonic/periodic and bursty edges, the results shown in Figures 7C-F suggest that between network integration is dominated by bursty inter-contact profiles rather than tonic/periodic contact profiles. By taking the average of the burstiness coefficient (*B, see eq. [2]*) for all between-RSN edges and within-RSN edges (where an edge is present in at least one state) demonstrates that the between-RSN edges have a bursty coefficient (on average) and the within-RSNs edges, on the other hand, have a tonic/periodic coefficient (on average, see Figures 7GH). For both edge thresholds (top 5 and 10 percent), there was a significant increase for the average between-RSN bursty coefficient compared to the average within-RSN bursty coefficient ($p<0.05$). The results presented in Figure 7 lends support to the view that the large-scale integration and the flow of information between resting-state networks are largely implemented in the form of bursts, i.e. an absence of connectivity for periods of time are followed by bursts of activity during which an exchange in information between resting-state networks occurs.

**Figure 7**

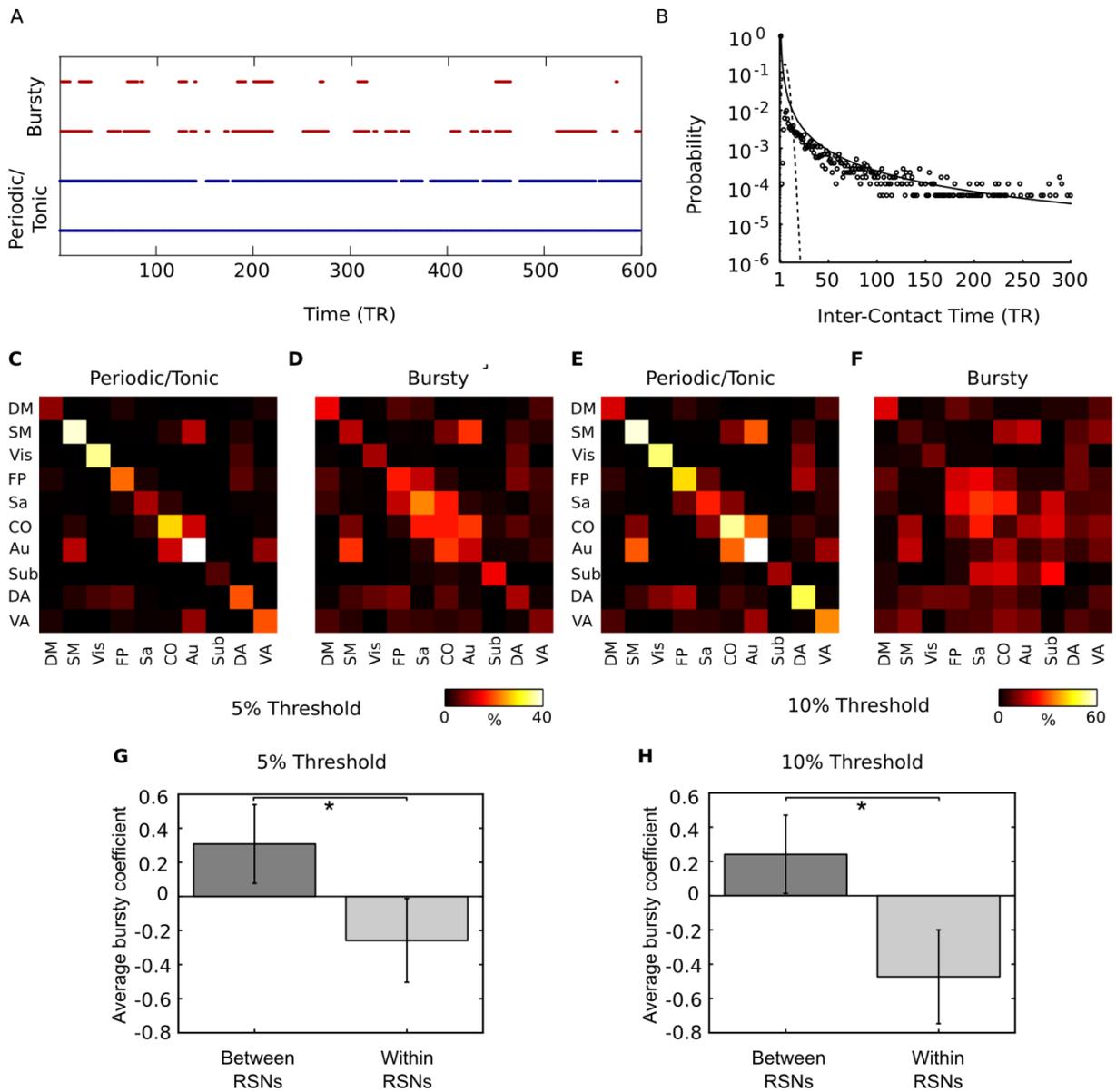

*Figure 7.* The burstiness of between- and within-resting-state brain connectivity. (A) An example of two bursty (red) and two periodic edges (blue) extracted from one representative subject while retaining the top 5 percent of the strongest edges (only the first half of the fMRI session is shown). Line indicates the presence of the edge. Note that periodic edges of connectivity can always be present. (B) An example of the distribution of inter-contact times for a single edge of brain connectivity. Data from all subjects were included and the graph was cropped to show the inter-contact times up to 300 TRs (216 seconds). The dashed line shows a fitted Poisson distribution and the solid line shows an *approximate Pareto distribution ($T^{-a}$) with $a$ set to 1.8 and T is the inter-contact times. The bursty nature of brain connectivity is evident by the fact that the distribution of inter-contact times displays a heavy fat-tail with a varying degree of longer inter-contact time as well as a cluster of very short inter-contact times. This behavior is in stark contrast to the corresponding Poisson distribution. (C) Percentage of edges between each combination of resting-state networks that had a periodic temporal pattern (p<0.05, two tailed). Note that the diagonal elements (within RSN-connectivity) had the majority of the periodic edges. (D) Percentage of edges between each RSN combination that had bursty edges (p<0.05, two tailed) when keeping the top 5% of all edges. Note that the between-RSN integration shows a large presence of bursty connectivity. (E) Same as C but keeping the top 10 % of all edges. (F) Same as D but keeping the top 10% of all edges. (G) Average burstiness coefficient (B) for all between-RSN edges and within-RSN edges when withholding the top 5% of all edges (edges must be present in at least one time-point) showing a statistical difference between the two (p<0.05). Error-bars show standard deviation. (H) Same as G but withholding the top 10% of all edges (p<0.05).*

A possible concern is whether burstiness was induced by the choice of cluster dimensionality in the k-means algorithm. This concern is also unwarranted, since a similar pattern of burstiness for between-network connectivity using the identical edge thresholds can be founds when k is set to 5 and 12 (see Supplementary Figures S3 and S7).

**Discussion**

We have presented a novel framework for dynamic functional connectivity that in part builds upon previous point-based methods [Liu et al., 2013; Tagliazucchi et al., 2012]. However, several features of the proposed method make it very suitable for the investigation of the temporal dynamics of resting-state fMRI data. Importantly, the presented method incorporates the whole time-series in the analysis and therefore takes advantage of exploiting the full range of dynamical changes in connectivity over time. We used a clustering analysis to create so called s-graphlets *(state-graphlets)* that captures the connectivity profiles of each individual state. In a subsequent step, we reuse the s-graphets to form a time-series of graphs (i.e. t-graphlets or *temporal-graphlets*) that allows for a detailed temporal analysis employing tools from temporal graph theory. It is worth mentioning that the proposed method does not need a temporal window length to be specified and it thereby avoids the problem of bias to certain signal frequencies that are inherent to sliding window techniques. The fact that each correlation matrix at all time-points is not dependent on its immediate neighbors allows for a greater temporal sensitivity in the analysis, which in turn provides the opportunity to follow fast transitions between states. Using the framework described in the present paper, we have presented two temporal graph properties in the brain.

*The burstiness and persistency of brain large-scale resting-state network connectivity.*

Interestingly, the conception of bursts of neuronal activity was founded a rather long time ago in the neuroscience literature and its origins can be traced back to studies that conducted electrophysiological recordings in single neurons (e.g. "complex spikes" in the hippocampus, Renck 1973). Moreover, the idea that bursts of brain activity may provide a vital mechanism for neuronal synchronization in the brain together with the notion that neuronal communication by means of bursts inherently provides richness in information content has gained wide-spread acceptance [Lisman 1997; Izhikevich 2000, 2006]. Importantly, previous research has shown that bursts of neuronal spiking are spontaneously occurring in the brain [Connors & Gutnick 1990], which suggests that bursts of brain activity can be monitored during resting-state conditions in the human brain. Thus, our results regarding the burstiness of brain connectivity indicate that the temporal burstiness of neuronal communication transcends levels of analysis and that it can be observed using functional neuroimaging techniques. However, we would like to emphasize the fact that our conceptualization of burstiness from the vantage point of large-scale brain network connectivity should not be interpreted categorically. That is, we do not in any way imply that bursts of brain connectivity as described in the present work can be directly traced from specific patterns of "bursty" activity at a level of single neuron recordings. While it would have been possible for us to describe our findings in more neutral terms such as e.g. "transients", which do not carry the same historical neuroscientific connotations as bursts, we strongly believe that there are good reasons to use the term bursts in the context of describing the dynamics of large-scale brain network connectivity, especially considering the acceptance of the usage of the term bursts in the temporal graph theoretical literature.

The key feature of a bursty mode of communication, regardless whether it applies to the temporal patterns of e-mail correspondence, the spread of an infectious disease, the firing patterns of single neurons or, as in our case, transient increases of large-scale brain network connectivity, is its properties in terms of the temporal evolution of points of contact (or connectivity) that is the focus of our attention. Namely, when single neurons display a bursting pattern of activity, they repeatedly fire discrete groups (or bursts) of spikes. Each group/burst of activity is followed by a period of quiescence before the next burst occurs. This temporal behavior of bursts can readily be expressed in statistical terms where the inter-contact times between bursts of activity, regardless of their physiological origin, are well approximated by a fat-tailed distribution. A fat-tailed distribution, (e.g. a Pareto distribution) is characterized by a strong presence of very short as well as quite long inter-contact times which makes it distinctly different from the distribution obtained from e.g. a Poisson process, a distribution that is often used to model discrete events in time.

To this end, the results shown Figure 7B strongly suggests that the temporal evolution of brain connectivity is in good agreement with the notion of a bursty mode of communication. Further, the results shown in Figures 7A-I and Supplementary Figure S7 provide further experimental evidence that a bursty mode of communication is indeed present between resting-state networks as well as for all states. The use of temporal graph theory combined with the point-based approach to investigate dynamic resting-state fMRI connectivity has allowed us to gain new insights into brain connectivity. In particular, we have introduced the concept of bursts in the study of large-scale brain connectivity and shown that they can be detected between the previously well characterized resting-state networks. However, the question whether the bursty activity identified in this work can be correlated with the underlying neuronal activity is still an open question that should be target in more detail in future research.

The results presented here using the high temporal sensitivity obtained by the PBM method gains further support from previous work that has reported the cognitive and behavioral significance of point-by-point changes in the BOLD signal. For example, using a sensory discrimination task in an event-related fMRI study, Boly and colleagues showed that pre-stimuli baseline fluctuations of the BOLD signal positively predicted brain responses in the medial thalamus and the fronto-parietal network, whereas a negative correlation between subsequent reporting of conscious perception and pre-stimuli processing in the posteromedial parietal cortex [Boly et al., 2007]. In a similar vein, it has been shown that spontaneous pre-stimuli BOLD brain activity preceding presentation of auditory near-threshold stimuli correlate with the detection rate [Sadaghiani et al., 2009]. These studies and others suggest that point-by-point changes in spontaneous BOLD activity are relevant for behavioral output. Thus, it seems plausible that the PBM method could provide the means to investigate integrative processes between separate brain networks at the level of individual sampled time-points and hence provide further insight on the issue of the relationship between spontaneous pre-stimuli baseline activity and behavioral output (see also recent review by Sadaghiani and Kleinschmidt, 2013).

Persistency relates to the "memory" of connectivity profiles in the brain, i.e. how long there still exists some similarity in brain connectivity with respect to a given point in time. Depending on the given task, this persistency measure may vary. If so, estimates of persistency of brain connectivity might have the potential to become a powerful metric. Tentatively, we envision that measures of persistency might prove to be useful when one is interested in how long a certain cognitive process or external event gives a footprint in the brain connectivity profile. However, we cannot at present rule out the possibility that persistency may simply represent the convolution of neuronal response with the hemodynamic response function in spontaneous non-event locked brain activity and does not vary in relation to e.g. cognitive load. Note that the persistency measure defined here does not need to be used as a global measure, but can equally well be estimated at the level of an individual edge or resting-state network. In future work, it would be worthwhile to investigate its sensitivity with regards to each state or to different networks.

The burstiness of between-RSN connectivity shown here provides a rationale for why previous studies on resting-state fMRI unequivocally have found spatially well segregated RSN but far less information regarding network integration, since its bursty nature makes it harder to detect in static connectivity studies. It seems likely that the point-based dynamic connectivity method has the potential to be applied in studies that try to isolate internal events that might be related to changes in awareness, sleepiness or other physiological changes during periods of rest. An interesting avenue of further research would be to investigate a putative link between bursts of brain connectivity and its relationship to the underlying neurophysiology. Although ample experimental evidence has been presented that suggest that the BOLD signal strongly correlate with the local field potential (LFP) [Logothetis et al., 2001], little is known of how the temporal dynamics of the BOLD signal interact with the different frequency bands within the LFP. However, a study has shown that the amplitude and temporal evolution of the BOLD signal relates to different frequencies of the LFP [Magri et al., 2012]. These recent findings might provide an experimental scaffolding to investigate the primary physiological mechanisms in terms of LFP frequency characteristics that might be invoked in bursts of brain network connectivity.

It is well known that the BOLD response is a neurovascular event that is rather sluggish in time and this might cast doubts on whether the rather fast transitions between states shown here are indeed feasible. However, it should be kept in mind that we have shown that there is a higher probability of a state transition to occur between states/s-graphlets that are closer to each other compared to states that are more distant. Furthermore, we have shown that for the eight s-graphlets utilized, the persistency in connectivity profiles is 13.7 seconds, a time that seems reasonable from the point of view that the sluggishness of the BOLD response is acting in a constraining manner on the rate of transitions between states. Other properties regarding the temporal dynamics and dynamic coordination of the brain have been proposed, one such property is metastability where coordination occurs through brief moments of stable coordination between the underlying dimensions [Tognoli and Kelso 2009, 2014]. Future work might want to consider the relationship between bursty connectivity and metastable states in the human brain.

*Methodological limitations.*

There are several limitation and considerations that needs to be discussed in relation to the proposed method and we will outline them below. First, we would like to emphasize the fact that any type of clustering algorithm can be used for the point-based method, but whichever clustering technique that is used, it bears its own disadvantages as well as advantages. We opted to use the k-means algorithm mainly due to its previous use on fMRI data and its simplicity. However, the problem of selecting an appropriate cluster dimensionality is a methodological concern that the k-means algorithm shares with most other clustering techniques. We believe that our choice of $k = 8$ is rather conservative and well in line with previous investigations of dynamic resting-state functional connectivity [Allen et al., 2014; Liu & Duyn 2013]. Further, since our overall findings of a predominant bursty mode of between-network connectivity together with a largely period/tonic mode of within-network communication were consistent for a wide range of choices of k (k = 5 and 12, see Supplementary Figure S7) as well as investigated in an independent dataset (see Supplementary Figure S3), we feel confident that our findings are not critically dependent on cluster dimensionality.

As mentioned throughout the paper, the creation of t-graphlets by the insertion of the s-graphlets into a time-series of connectivity matrices puts limits on the achieved overall variance in connectivity, and it is bounded by the *k* number of discrete possible states. Two questions come to mind from this limitation. (i) is the limited set of connectivity estimates relevant and (ii) is the mapping from s-graphlets to t-graphlets reasonable? We would like to address the first questions by stating that the connectivity estimates presented in the present work are fully compatible with the general underlying assumptions of functional connectivity that posit that the interaction between brain regions can inferred from the their degree of covariance. By splitting the entire dataset into clusters in which the global context of brain connectivity (i.e. the ROIs relative spatial activity amongst all other nodes) is different from each other, allowed us to infer the variability of interaction that was based on the covariance when the global spatial pattern is in a given state as defined by the clustering. The second concern regards whether the mapping from s-graphlet to t-graphlet is reasonable. What we have presented here is a simple mapping procedure to illustrate the conceptual idea of the PBM framework, but more complex methods of creating t-graphlets are certainly possible. This can be addressed in the future in numerous ways. A possible strategy to cope with this problem would be to take into account all the individual distances in state-space between all cluster centroids when computing the t-graphlets. Thus, the suggested strategy would essentially result in that weighted t-graphlets are created so that each time-point has a unique corresponding connectivity matrix that reflects the relative distances to all states at each point in time. However, the weighting schemes outlined above are not without their own inherent assumptions. Work to address this issue by developing a weighted version of the PBM will be presented in a separate paper.

Another concern that deserves to be mentioned is that we calculated measures of burstiness on a magnitude-based thresholding of t-graphlets. We noted that, with this approach, the within-network connectivity is predominately tonic and/or periodic. It has been noted using other methods of dynamic functional connectivity that the variance of within-network connectivity is smaller than the between-network connectivity [Thompson & Fransson 2015b]. Thus, it cannot be ruled out that the within-network connectivity also to some degree take place in the form of bursts but has, on the whole, a larger degree of sustained connectivity. This issue can in the future be investigated more thoroughly when a weighted t-graphlet metric has been developed and thresholds for each time-series might be a more appropriate strategy than the global proportional threshold applied here. Finally, at present the observed burstiness of large-scale network connectivity patterns has only been observed in the BOLD signal. Until replicated using magnetoencephalography or simultaneous-electroencephalogram and fMRI recordings, one cannot completely rule out the possibility that the properties reported here are merely property of the BOLD signal and not a neuronal property.

*Conclusion.*

We have presented a novel point-based method (PBM) that takes full advantage of the resolution provided by low TR resting-state fMRI. In doing so we have shown that measures derived from the theory of temporal graphs can be successfully applied to resting-state fMRI data. This has resulted in new insights regarding how functional integration between previously well-known, segregated resting-state networks occurs in time. The results from the point-based dynamic functional connectivity method suggests that the functional integration between resting-state networks has a unique temporal evolution during resting-state conditions that encompasses short bursty periods of connectivity that are dispersed by intermittent periods characterized by a low degree of between-network connectivity. In contrast, within-resting-state network connectivity displays both periodic and bursty patterns of connectivity changes over time. Moreover, the point-based approach to study the dynamical properties of functional connectivity allowed us to estimate the duration of the intrinsic trace or memory of established connectivity patterns in the human brain. We believe that the presented method holds promise to be used as a versatile tool to quantify dynamic network integration between segregated networks during different experimental conditions.

**Acknowledgements:** P.F. was supported by the Swedish Research Council (grant no. 621-2012-4911). Data were provided by the Human Connectome Project, WU-Minn Consortium (Principal Investigators: David Van Essen and Kamil Ugurbil; 1U54MH091657) funded by the 16 NIH Institutes and Centers that support the NIH Blueprint for Neuroscience Research; and by the McDonnell Center for Systems Neuroscience at Washington University. The funders had no role in study design, data collection and analysis, decision to publish, or preparation of the manuscript.

**References**

Allan TW, Francis ST, Caballero-Gaudes C, Morris PG, Liddle EB, Liddle PF, Brookes MJ, Gowland PA (2015): Functional Connectivity in MRI Is Driven by Spontaneous BOLD Events. PLoS One 10:e0124577.

Allen EA, Damaraju E, Plis SM, Erhardt EB, Eichele T, Calhoun VD (2014): Tracking whole-brain connectivity dynamics in the resting state. Cereb Cortex 24:663–676.

Barabási AL (2005): The origin of bursts and heavy tails in human dynamics. Nature 435: 207–211.

Boly M, Balteau E, Schnakers C, Degueldre C, Moonen G, Luxen A, Philips A, Peigneux P, Maquet P, Laureys S (2007): Baseline brain activity fluctuations predict somatosensory perception in humans. Proc Natl Acad Sci USA 104:12187-12192.

Bullmore E, Sporns O (2009): Complex brain networks: graph theoretical analysis of structural and functional systems. Nat Rev Neurosci 10:186–198.

Cole MW, Reynolds JR, Power JD, Repovs G, Anticevic A, Braver RS (2013): Multi-task connectivity reveals flexible hubs for adaptive task control. Nat Neurosci 16:1348–1355.

Connors BW, Gutnick MJ (1990): Intrinsic firing patterns of diverse neocortical neurons. Trends Neurosci 13:99-104.

Damoiseaux JS, Rombouts SA, Barkhof F, Scheltens P, Stam CJ, Smith SM, Beckmann CF (2006): Consistent resting-state networks across healthy subjects. Proc Natl Acad Sci USA 103:13848–13853.

Davis B, Jovicich J, Iacovella V, Hasson U (2013): Functional and developmental significance of amplitude variance asymmetry in the BOLD resting-state signal. Cereb Cortex 24:1332–1350.

De Luca M, Beckmann CF, De Stefano N, Matthews PM, Smith M (2006): fMRI resting state networks define distinct modes of long-distance interactions in the human brain. NeuroImage 29:1359–1367.

Fox MD, Snyder AZ, Vincent JL, Corbetta M, van Essen DC, Raichle ME (2005): The human brain is intrinsically organized into dynamic, anticorrelated functional networks. Proc Natl Acad Sci USA 102:9673–9678.

Fransson P (2005): Spontaneous low-frequency BOLD signal fluctuations: an fMRI investigation of the resting-state default mode of brain function hypothesis. Hum Brain Mapp 26:15–29.

Glasser MF, Sotiropoulos SN, Wilson JA, Coalson TS, Fischl B, Andersson J, et al. (2013): The minimal preprocessing pipelines for the Human Connectome Project. NeuroImage 80:105–124.

Goh KI, Barabási AL (2008): Burstiness and Memory in Complex Systems. arXiv:physics/0610233.

Greicius MD, Krasnow B, Reiss AL, Menon V (2003): Functional connectivity in the resting brain: A network analysis of the default mode of brain function. Proc Natl Acad Sci USA 100:253–258.

Griffanti L, Salami-Khorshidi G, Beckmann CF, Auerbach EJ, Douaud G, Sexton CE, et al. (2014): ICA-based artifact removal and accelerated fMRI acquisition for improved resting-state network imaging. NeuroImage 95:232-247.

Holme P, Saramäki J (2012): Temporal networks. Physics Rep 519:97–125.

Hutchison, R.M., Womelsdorf T, Allen EA, Bandettini PA, Calhoun VD, Corbetta M et al. (2013a): Dynamic functional connectivity: promise, issues, and interpretations. NeuroImage 80:360–378.

Hutchison RM, Womelsdorf T, Gati JS, Everling S, Menon RS (2013b): Resting-state networks show dynamic functional connectivity in awake humans and anesthetized macaques. Hum Brain Mapp 34:2154–2177.

Izhikevich EM (2000): Neural excitability, spiking and bursting. Int J Bifurc Chaos 10: 1171-1266.

Izhikevich EM (2006): Bursting. Scholarpedia, 1(3):1300.

Karsai M, Kivelä M, Pan RK, Kaski K, Kertesz J, Barabási AL, Saramäki J (2011): Small but slow world: how network topology and burstiness slow down spreading. Phys Rev E Stat Nonlin Soft Matter Phys 83:025102.

Keilholz SD, Magnuson ME, Pan WJ, Willis M, Thompson GJ (2013): Dynamic properties of functional connectivity in the rodent. Brain Conn 3:31-40.

Kiviniemi V, Vire T, Remes J, Elseoud AA, Starck T, Tervonen O, Nikkinen JA (2011): Sliding time-window ICA reveals spatial variability of the default mode network in time. Brain Conn 1:339–347.

Kuhn HW (1955): The Hungarian method for the assignment problem. Naval research logistics quarterly 2(1-2):83-97.

Leonardi N, Richiardi J, Gschwind M, Simioni S, Annoni JM, Schluep M, et al. (2013): Principal components of functional connectivity: a new approach to study dynamic brain connectivity during rest. NeuroImage 83:937–950.

Lisman JE (1997): Bursts as a unit of neural information: making unreliable synapses reliable. Trends Neurosci 20:38-43.

Liu X, Duyn JH (2013): Time-varying functional network information extracted from brief instances of spontaneous brain activity. Proc Natl Acad Sci USA 110:4392–4397.

Logothetis NK, Pauls J, Augath M, Trinath T, Oeltermann A (2001): Neurophysiological investigation of the basis of the fMRI signal. Nature 412:150-157.

Magri C, Schridde U, Murayama Y, Panzeri S, Logothetis NK (2012): The amplitude and timing of the BOLD signal reflects the relationship between local field potential power at different frequencies. J Neurosci 32:1395-1407.

Majeed W, Magnuson M, Hasenkamp W, Schwarb W, Schumacher EH, Barsalou L, Keilholz SD (2011): Spatiotemporal dynamics of low frequency BOLD fluctuations in rats and humans. NeuroImage 54:1140-1150.

Munkres J (1957): Algorithms for the assignment and transportation problems. Journal of the Society for Industrial and Applied Mathematics 5(1):32-38.

Nicosia V et al. (2013): Graph Metrics for Temporal Networks. In Temporal Networks. Berlin Heidelberg: Springer; 2013 pp. 15–40.

Power JD, Cohen AL, Nelson SM, Wig GS, Barnes KA, Church JA, et al. (2011): Functional network organization of the human brain. Neuron 72:665–678.

Power JD, Barnes K, Snyder AZ, Schlaggar BL, Petersen SE (2012): Spurious but systematic correlations in functional connectivity MRI networks arise from subject motion. NeuroImage 59:2142–2154.

Renck JB (1973): Studies on single neurons in dorsal hippocampal formation and septum in unrestrained rats. Exp Neurol 41:461-533.

Sadaghiani S, Hesselmann G, Kleinschmidt A (2009): Distributed and antagonistic contributions of ongoing activity fluctuations to auditory stimulus detection. J Neurosci 29:13410-13417.

Sadaghiani S, Kleinschmidt A (2013): Functional interactions between intrinsic brain activity and behavior. NeuroImage 80:379-386.

Salimi-Khorshidi G, Douaud G, Beckmann CF, Glass MF, Griffanti L, Smith SM (2014): Automatic denoising of functional MRI data: combining independent component analysis and hierarchical fusion of classifiers. NeuroImage 90:449-468.

Smith SM, Beckmann CF, Andersson J, Auerbach EJ, Bijsterbosch J, Douaud G et al. (2013): Resting-state fMRI in the Human Connectome Project. NeuroImage 80:144–168.

Smith SM, Miller KL, Moeller S, Xu J, Auerbach EJ, Woolrich MW, et al. (2012): Temporally-independent functional modes of spontaneous brain activity. Proc Natl Acad Sci USA 109:3131–3136.

Tagliazucchi E, Balenzuela P, Fraiman D, Montoya P, Chialvo DR (2012): Criticality in large-scale brain FMRI dynamics unveiled by a novel point process analysis. Front Physiol 3:15.

Takaguchi T, Masuda N, Holme P (2013): Bursty communication patterns facilitate spreading in a threshold-based epidemic dynamics. PLoS One. 8:p.e68629.

Thompson GA, Merrit MD, Pan WJ, Magnuson ME, Grooms JK, Jaeger D, Keilholz SD (2013): Neural correlates of time-varying functional connectivity in the rat. NeuroImage, 83:826-836.

Thompson WH, Fransson P (2015a): The frequency dimension of fMRI dynamic connectivity: network connectivity, functional hubs and integration in the resting brain. Neuroimage 121:227-242.

Thompson WH, Fransson P (2015b): The mean–variance relationship reveals two possible strategies for dynamic brain connectivity analysis in fMRI. Frontiers Hum Neurosci 9:1-7.

Tognoli E, Kelso JA (2009): Brain coordination dynamics: true and false faces of phase synchrony and metastability. Prog Neurobiol 87:31–40.

Tognoli E, Kelso, JA (2014): The metastable brain. Neuron 81, 35–48.

Ugurbil K, Xu J, Auerbach EJ, Moeller S, Vu AT, Duarte-Carvajalina JM (2013): Pushing the spatial and temporal resolution for functional and diffusion MRI in the Human Connectome Project. NeuroImage 80:80-104.

Van Dijk KR, Sabuncu MR, Buckner RL (2012): The influence of head motion on intrinsic functional connectivity MRI. NeuroImage 59:431–438.

Van Essen DC, Ugurbil K, Auerbach EJ, Barch D, Behrens TE, Bucholz R et al. (2012): The Human Connectome Project: a data acquisition perspective. NeuroImage 62:2222–2231.

Van Essen DC, Smith SM, Barch DM, Behrens TEJ, Yacoub E, Ugurbil K et al. (2013): The WU-Minn Human Connectome Project: An overview. NeuroImage 80:62–79.

**Supplementary Figures.**

**Figure S1**

A

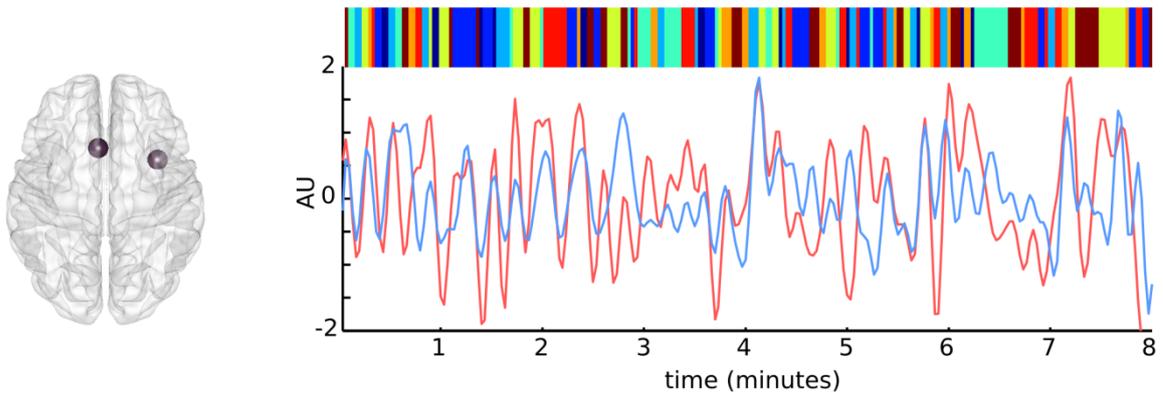

B

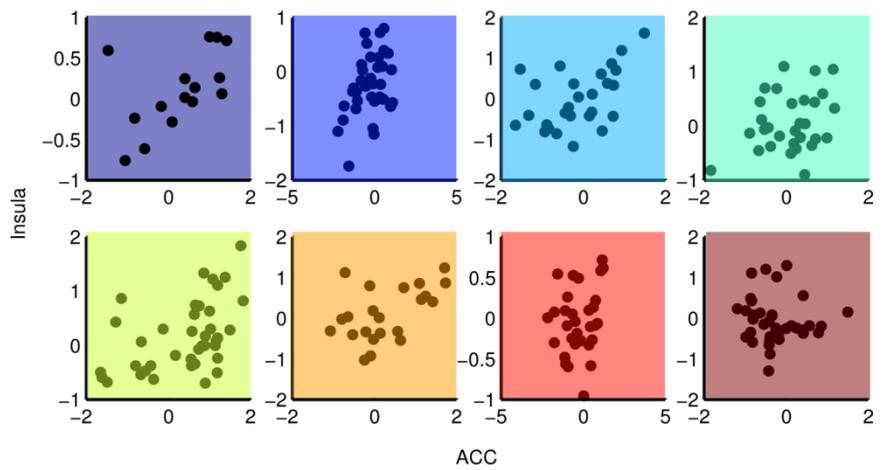

C

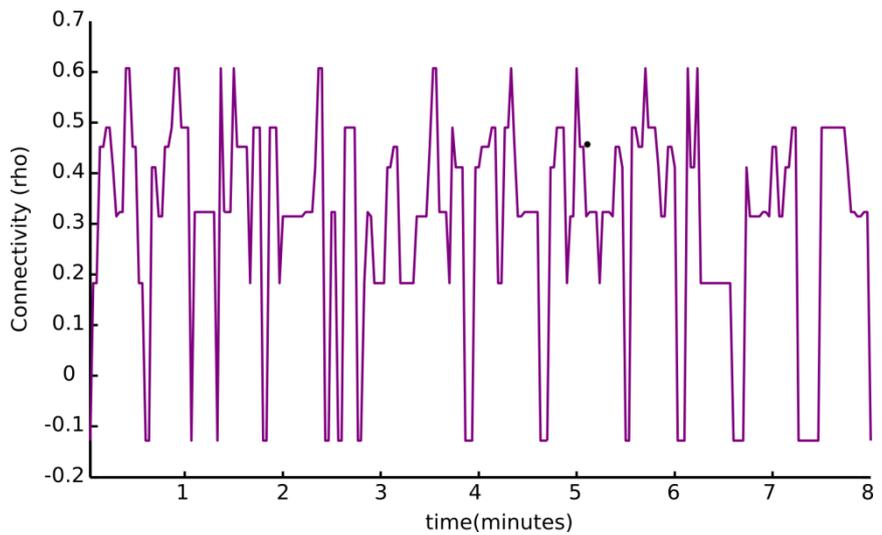

*Figure S1.* An illustration of the concept of estimating the co-variance between ROI signal intensity time-courses based on non-neighboring time-points in different clusters and the creation of functional connectivity time-series. In this example, the BOLD signal

*intensity time courses from two seed regions located in the ACC (red line) and insula (blue line) are first extracted (A). For simplicity, we here show for a single subject and, in contrast to Figure 1, how the PBM method effects a single connection between two brain regions, rather than the entire graph. The signal intensity time-series are subsequently divided into different clusters using the k-means clustering algorithm (see Figure 1 and Methods section) which are here shown for the case of k=8 using a coloring scheme displayed above the signal intensity time-series in panel (A). In a subsequent step, the co-variance is computed for all time-points that are allocated for each cluster (B). The final step is shown in panel (C) where a functional connectivity time-series between the ACC and the insula is created based on the co-variance computed in (B) for the eight different clusters. An additional descriptive illustration that includes the clustering step is given in Figure 1 and a detailed account is given in the methods section.*

*Figure S2*

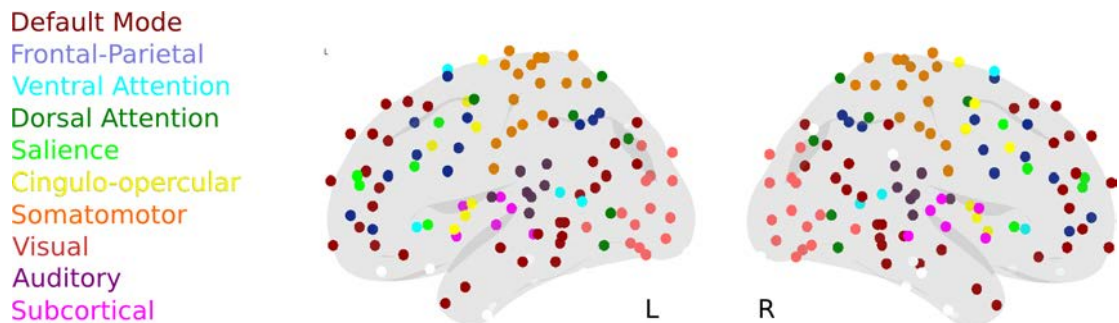

*Figure S2. Regions of interest and the resting state network template chosen. 264 regions of interest (10mm spheres) was defined (Power et al. 2011) and resting state network assignments based on (Power et al., 2011 and Cole et al., 2013) for each node corresponding to the color of the text. Unclassified nodes are shown in white. Cerebellum is not shown. These regions and network assignments are used through the analysis.*

**Figure S3**

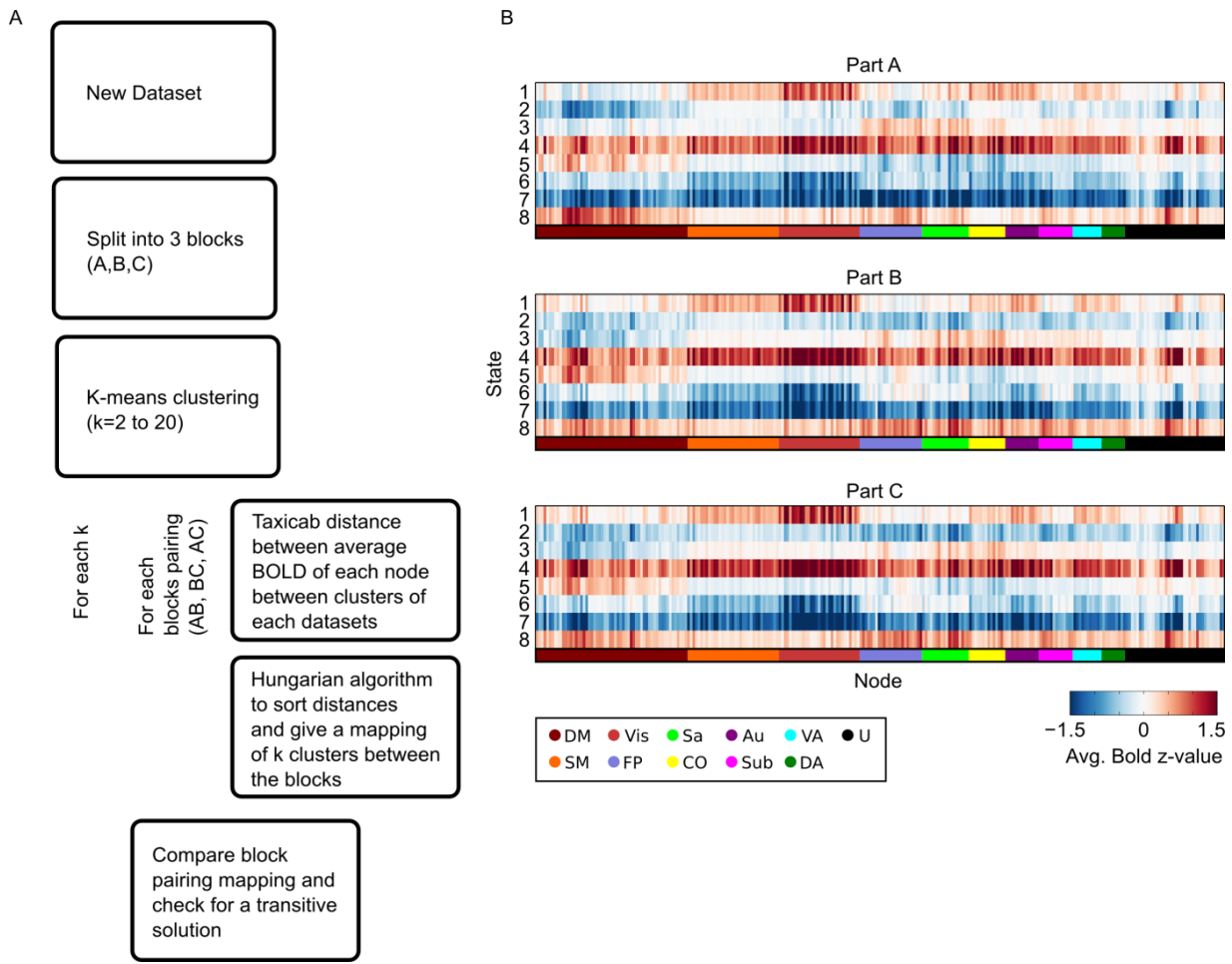

*Figure S3.* This figure shows a schematic plot together with example data that visualizes the strategy used to further justifying our choice of k in a separate independent dataset. (A) Shows a schematic outline of the work-flow described in the methods section. Panel (B) shows the average BOLD standardized intensity across time for all nodes and states in three different blocks (A, B and C) of the second dataset when k=8. The clusters were sorted using the Hungarian algorithm (see also for comparison Figure 2C and the methods section for further details).

**Figure S4**

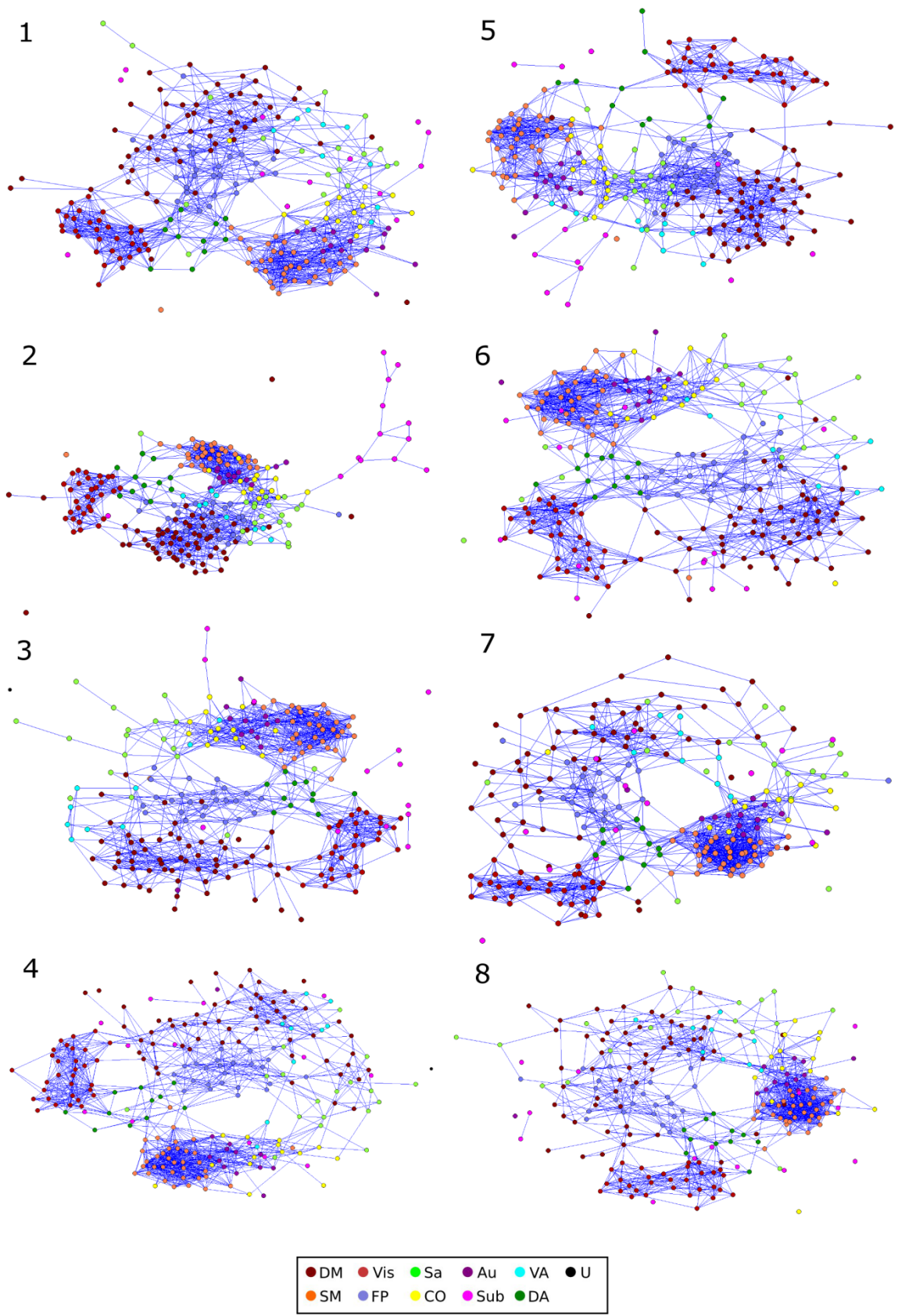

*Figure S4.* The eight states/s-graphlets derived from the k-means clustering are here shown using the spring-embedded Kamada-Kawai representation. Only the top 5 percent of all edges, similar to Figure 2D, were included in the plots.

*Figure S5*

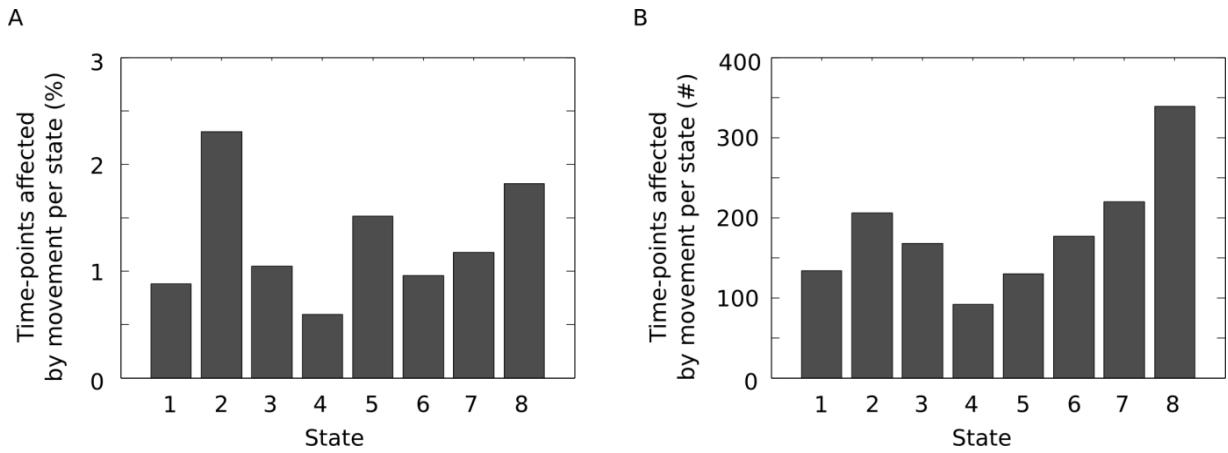

*Figure S5.* No state corresponds exclusively to movement. Distribution of time-points assigned to each state at time-points for which FD>0.5, indicating high micro-movements. There is no state which is exclusively found to be associated with estimated micro-movement artifacts.

*Figure S6*

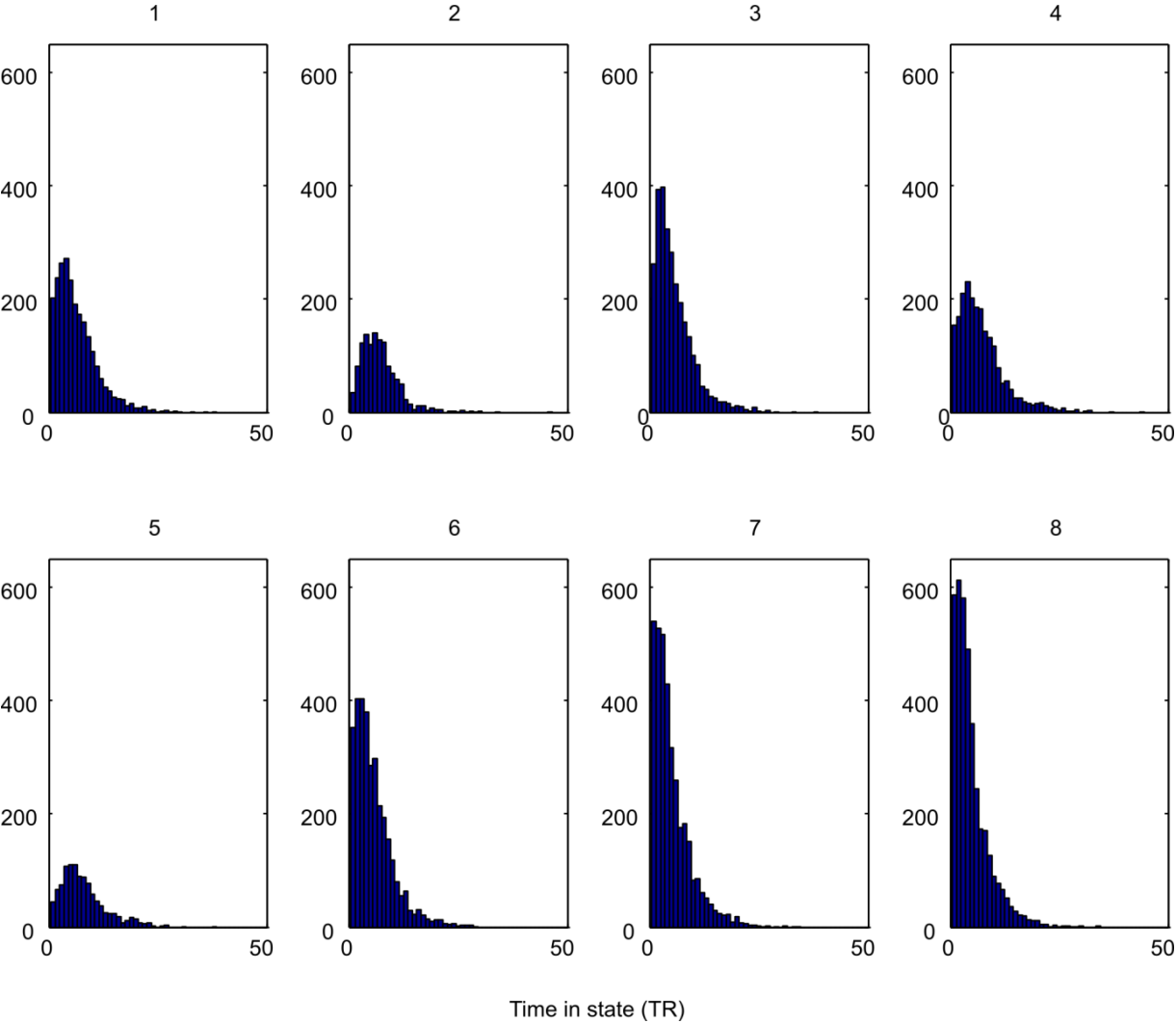

*Figure S6.* Histograms showing the duration of stay in all eight states before a switch to another state occurred. The number of state is shown at the top of each histogram.

*Figure S7*

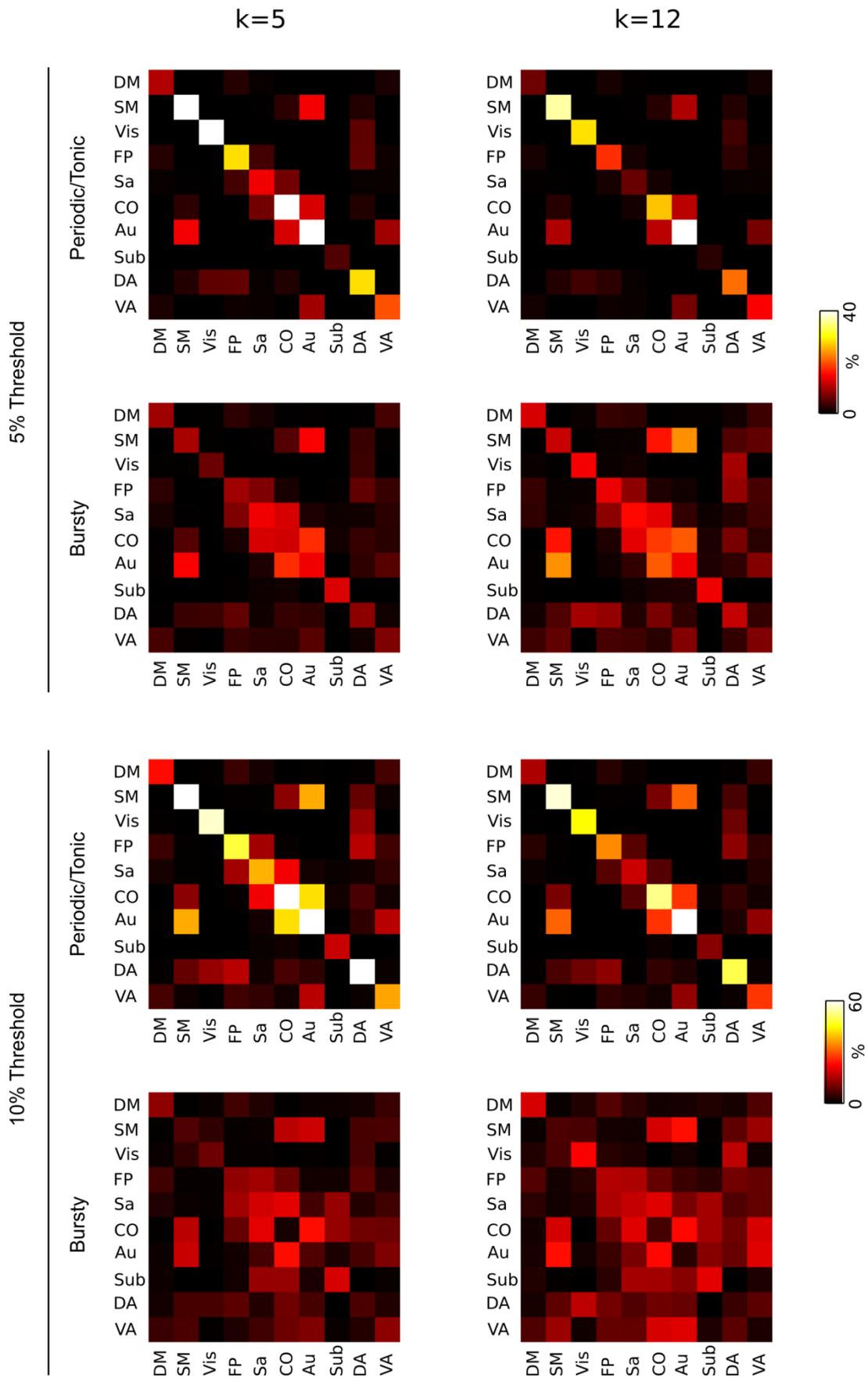

*Figure S7.* Replication of the burstiness of brain connectivity for different choices of k (k = 5 and 12 in the k-means algorithm) used to derive s-graphlets. Similar to the results shown in Figs. 7C-F, the percentage of edges that were significantly bursty or tonic/periodic ($p \leq 0.05$, two-tailed) is shown using two different edge thresholds (withholding either the top 5 or 10 percent of all edges).